\documentclass[letterpaper,preprintnumbers,twocolumn,eqsecnum,superscriptaddress,aps,nofootinbib,]{revtex4}
\usepackage{amssymb}\usepackage[centertags]{amsmath}\usepackage{txfonts}\usepackage{epsfig}\usepackage{bm}
\usepackage{color}\usepackage{graphicx,graphics}\usepackage{multirow}\usepackage{float}\usepackage{ulem}
\usepackage{hyperref}\usepackage{setspace}\usepackage{slashed}\usepackage{booktabs}
\hypersetup{colorlinks=true, citecolor=blue, linkcolor=red, filecolor=black,urlcolor=blue}
\allowdisplaybreaks[4]

\begin{document}

\title{The semi-inclusive deeply inelastic scattering  in the $eN$ collinear frame}


\author{Weihua Yang}
\affiliation{College of Nuclear Equipment and Nuclear Engineering, Yantai University,\\ Yantai, Shandong 264005, China}


\begin{abstract}
The deeply inelastic scattering is one of the most important processes in studying the nucleon structure. Theoretical calculations for both the inclusive one and the semi-inclusive one are generally carried out in the virtual photon-nucleon collinear frame in which virtual photon does not have the transverse components. Expressions in this frame are written in relatively simple forms. Nevertheless, it is also meaningful to calculate the scattering process in the electron-nucleon collinear frame where new measurement schemes are obtained. In the present  paper, we reconsider the semi-inclusive deeply inelastic scattering process in the electron-nucleon collinear frame and present the results of azimuthal asymmetries and quark intrinsic asymmetries. We find that the differential cross sections in  these two frames are the same at leading twist level but different at higher twist level. Azimuthal asymmetries and intrinsic asymmetries in these two frames have the same forms but different kinematic factors. For the sake of completeness, both the electromagnetic and weak interactions are considered in our calculations. The neutral current measurements in the scattering process could be used as electroweak precision tests which can provide new accurate determinations of the electroweak couplings.

\end{abstract}

\maketitle

\section{Introduction}

The deeply inelastic scattering (DIS) experiments provided a unique window in studying the nucleon structure  in the past decades via the lepton-nucleon reactions. It will still play an important role in the future Electron-Ion Collider (EIC)~\cite{Accardi:2012qut,AbdulKhalek:2021gbh,Anderle:2021wcy} experiments.  The inclusive DIS process  can only access the longitudinal motion of partons in a fast moving nucleon or the longitudinal momentum distributions along the light-cone direction determined by the nucleon.  In order to resolve the  transverse momentum distributions, one is supposed to consider the semi-inclusive DIS (SIDIS) where a final state hadron or a  jet  is also measured in addition to the scattered lepton. Under the one-photon exchange approximation, theoretical calculation are generally carried out in the virtual photon-nucleon ($\gamma^*N$) collinear frame in which virtual photon does not have the transverse components.  Systematic calculations for the hadron-production SIDIS process at leading order twist-3 level can be found in Refs.  \cite{Mulders:1995dh,Bacchetta:2006tn}. Further discussions at next-to-leading power can be found in Refs. \cite{Beneke:2022obx,Gamberg:2022lju,Rodini:2023plb}. In these reactions, the transverse momentum dependent  parton distribution functions (TMD PDFs) and fragmentation functions (FFs) have to be considered simultaneously. Uncertainties from TMD FFs  are inevitably introduced in extracting TMD PDFs. To avoid this problem, one considers the  jet-production SIDIS process where jets are generally taken as fermions (quarks) in the calculation. Here the jet and the scattered lepton are measured simultaneously. Comparing to the production of hadrons in the SIDIS process, the production of jets in a reaction takes on simpler forms that allows to calculate higher twist effects. Nevertheless, there is a shortcoming of this process  that can not be used to explore the chiral-odd quantities, e.g., Boer-Mulders function ($h_1^\perp$) \cite{Boer:1997nt}. Because no spin flip occurs in such a reaction that only the chiral-even contributions exist.  However, proposals for exploring the chiral-odd quantities through the jet fragmentation functions have been studied recently in Refs. \cite{Accardi:2017pmi,Accardi:2019luo,Accardi:2020iqn,Kang:2020xyq}.

Calculations for the jet-production SIDIS process in the $\gamma^*N$ collinear frame had been studied extensively \cite{Liang:2006wp,Song:2010pf,Song:2013sja,Wei:2016far,Gutierrez-Reyes:2018qez,Gutierrez-Reyes:2019vbx, Kang:2020fka,Arratia:2020ssx,Chen:2020ugq,Yang:2020qsk,Yang:2022xwy}.
Recently, the jet-production SIDIS process considered in the electron-nucleon ($e N$) collinear attracted much attentions \cite{Liu:2018trl,Liu:2020dct,Kang:2021ffh,Kang:2022dpx,H1:2021wkz}. A number of quantities were reconsidered, such as single spin asymmetry \cite{Liu:2018trl,Liu:2020dct,Kang:2021ffh,Kang:2022dpx}, parity violating asymmetries and charge asymmetries \cite{Yang:2023vyv}. Previous discussions of the jet-production SIDIS process in the $eN$ collinear frame were limited at leading twist level.
In this paper,  we reconsider this process and extend the calculation to the twist-3 level. This is not a naive extension because the current conversation law of the hadronic tensor ($q_\mu W^{\mu\nu}=0$) should be dealt with carefully. In the $\gamma^* N$ collinear frame, the current conservation at twist-3 is satisfied by the relationship  $q\cdot \bar q=q\cdot (q+2xp)=0$, see Ref. \cite{Chen:2020ugq}.  However, in the $e N$ collinear frame, the virtual-photon gains the transverse momentum component $q_T$. The previous relationship no longer holds, instead, a more complicated relationship is obtained.

In addition to the electromagnetic contributions, the (SI)DIS process also receives contributions from the weak interaction  from the exchange of the $Z^0$ boson. According to our numerical estimates, weak contributions will reach a few percent when $Q>10$ GeV in the (SI)DIS process.  The neutral current measurements in the scattering process could be used as electroweak precision tests \cite{Cahn:1977uu}  which can provide new accurate determinations of the electroweak couplings or new signals beyond standard model.  Furthermore, the charged current measurements can be used  in the flavor decomposition in PDF global analyses, specially for the determination of the parton distributions of the strange quark. In this paper, we limit ourselves by only calculate the neutral current SIDIS process in the $eN$ collinear frame.
After obtain the hadronic  tensor, we calculate the differential cross section, azimuthal asymmetries and intrinsic asymmetries \cite{Yang:2022xwy}. We find that the differential cross sections in  these two frames are the same in form at leading twist level but different at higher twist level. Azimuthal asymmetries and intrinsic asymmetries in these two frames have the same forms but different kinematic factors.

To be explicit, we organize this paper as follows. In Sec. \ref{sec:generalcs}, we present the formalism of the jet-production SIDIS process. Conventions used in this paper are also given. In Sec. \ref{sec:hadront}, we calculate the hadronic tensor up to twist-3 level in the $eN$ collinear frame. In Sec. \ref{sec:results} we present our calculation results, including differential cross section, azimuthal asymmetries and intrinsic asymmetries. Numerical estimates of the intrinsic asymmetry are also presented.  Finally, a brief summary is given in Sec. \ref{sec:summary}.

\section{The formalism} \label{sec:generalcs}

As mentioned in the introduction, we consider both the EM and weak interactions in the jet-production SIDIS.  Therefore, the exchange of a $Z^0$ boson between the electron and the nucleon can be relevant.
We label this reaction  in the following form,
\begin{align}
e^-(l,\lambda_e) + N(p,S) \rightarrow e^-(l^\prime) + q(k^\prime) + X,
\end{align}
where $\lambda_e$ is the helicity of the electron with momentum $l$, $N$ is a nucleon with momentum $p$ and spin 1/2. $q$ denotes a quark which corresponds to a jet of hadrons observed in experiments. We require that the jet is in the current fragmentation region.

The differential cross section of the SIDIS can be written as a product of the leptonic tensor and the hadronic tensor,
\begin{align}
  d\sigma = \frac{\alpha_{\rm em}^2}{sQ^4}A_r L^r_{\mu\nu}(l,\lambda_e, l^\prime)W_r^{\mu\nu}(q,p,S,k^\prime)\frac{d^3 l^\prime d^3 k}{E_{l^\prime}(2\pi)^22 E_{k^\prime}}. \label{f:crosssec}
\end{align}
The symbol $r$ can be $\gamma\gamma$, $ZZ$ and $\gamma Z$, for EM, weak and interference terms, respectively.
A summation over $r$ in Eq. (\ref{f:crosssec}) is understood, i.e., the total cross section is given by
\begin{align}
  d\sigma = d\sigma^{ZZ} + d\sigma^{\gamma Z} + d\sigma^{\gamma\gamma}.
\end{align}
$A_r$'s are defined as
\begin{align}
& A_{\gamma\gamma} = e_q^2, \nonumber\\
& A_{ZZ} = \frac{Q^4}{\left[(Q^2+M_Z^2)^2 + \Gamma_Z^2 M_Z^2 \right] \sin^4 2\theta_W} \equiv \chi, \nonumber\\
& A_{\gamma Z} = \frac{-2 e_q Q^2 (Q^2+M_Z^2)}{\left[(Q^2+M_Z^2)^2 + \Gamma_Z^2 M_Z^2 \right] \sin^2 2\theta_W}\equiv \chi_{int},
\end{align}
where $e_q$ is the electric charge of quark $q$. $M_Z, \Gamma_Z$ are the mass and width of $Z^0$ boson, $\theta_W$ is the weak mixing angle. $Q^2=-q^2=-(l-l')^2$. The leptonic tensors are respectively given by
\begin{align}
 &L^{\gamma\gamma}_{\mu\nu}(l,\lambda_e, l^\prime)= 2\left[ l_\mu l^\prime_\nu + l_\nu l^\prime_\mu - (l\cdot l^\prime)g_{\mu\nu}  \right] + 2i\lambda_e \varepsilon_{\mu\nu l l^\prime}, \label{f:leptongg}\\
 & L^{\gamma Z}_{\mu\nu}(l,\lambda_e, l^\prime)=(c_V^e - c_A^e \lambda_e) L^{\gamma\gamma}_{\mu\nu}(l,\lambda_e, l^\prime),\label{f:leptongz}\\
 & L^{ZZ}_{\mu\nu}(l,\lambda_e, l^\prime) =(c_1^e - c_3^e \lambda_e)L^{\gamma\gamma}_{\mu\nu}(l,\lambda_e, l^\prime), \label{f:leptonzz}
\end{align}
where $\lambda_e$ is the helicity of the electron. $c_1^e = (c_V^e)^2 + (c_A^e)^2$ and $c_3^e = 2 c_V^e c_A^e$.
$c_V^e$ and $c_A^e$ are defined in the weak interaction current $J^\mu (x)=\bar \psi(x)\Gamma^\mu\psi(x)$ with $\Gamma^\mu=\gamma^\mu(c_V^e-c_A^e\gamma^5)$.

The hadronic tensors are given by
\begin{align}
  W_{\gamma\gamma}^{\mu\nu}(q,p,k^\prime)& = \sum_X (2\pi)^3\delta^4(p + q - k^\prime- p_X) \nonumber \\
  &\times\langle p,S |J_{\gamma\gamma}^\mu(0)|k^\prime;X\rangle \langle k^\prime;X |J_{\gamma\gamma}^\nu(0)| p,S \rangle ,
  \label{f:wgg} \\
 W_{\gamma Z}^{\mu\nu}(q,p,k^\prime) &= \sum_X (2\pi)^3 \delta^4(p + q - k^\prime - p_X)\nonumber \\
  &\times\langle p,S |J_{ZZ}^\mu(0)|k^\prime;X\rangle \langle k^\prime;X | J_{\gamma\gamma}^\nu(0)| p,S \rangle ,
 \label{f:wgz} \\
  W_{ZZ}^{\mu\nu}(q,p,k^\prime)& = \sum_X  (2\pi)^3\delta^4(p + q - k^\prime - p_X)\nonumber \\
  &\times\langle p,S | J_{ZZ}^\mu(0)|k^\prime;X\rangle \langle k^\prime;X | J_{ZZ}^\nu(0) | p,S \rangle , \label{f:gzz}
\end{align}
where $J_{\gamma\gamma}^\mu(0) =\bar\psi(0) \gamma^\mu \psi(0)$, $ J_{ZZ}^\mu(0) =\bar\psi(0) \Gamma^\mu_q \psi(0)$ with $\Gamma^\mu_q = \gamma^\mu(c_V^q - c_A^q \gamma^5)$.
It is convenient to consider the  $k_T^\prime$-dependent cross section, i.e.,
\begin{align}
  d\sigma = \frac{\alpha_{\rm{em}}^2}{sQ^4}A_{r} L^r_{\mu\nu}(l,\lambda_e, l^\prime)W_r^{\mu\nu}(q,p,S,k_T^\prime) \frac{d^3 l^\prime d^2 k_T^\prime}{E_{l^\prime}}, \label{f:crosssection}
\end{align}
where the $k^{\prime}_z$ integrated  hadronic tensor is given by
\begin{align}
W_r^{\mu\nu}(q,p,S,k_T^\prime) = \int \frac{dk_z^\prime}{(2\pi)^32E_{k^\prime}} W_{r}^{\mu\nu}(q,p,S,k^\prime). \label{f:hadronzz}
\end{align}
With the convention of exploring the lepton-jet correlation \cite{Liu:2020dct}, we define $j=l^\prime+k^\prime$, the sum of the momenta of the scattered lepton and the final jet. In the $eN$ collinear frame,
\begin{align}
 \vec j_T = \vec l_T^\prime + \vec k_T^\prime =  \vec l_T^\prime + \vec k_T + \vec q_T =\vec k_T \label{f:jetk}
\end{align}
if the higher order radiation of gluon is neglected. In other words, the transverse momentum $\vec j_T$ equals to the intrinsic transverse momentum $\vec k_T$ of a quark in the nucleon which can induce the intrinsic asymmetry  \cite{Yang:2022xwy}. Therefore, the hadronic tensor and/or cross section can be defined in terms of the new momentum $j$ \cite{Yang:2023vyv}, i.e.,
\begin{align}
\frac{d\sigma}{d\eta d^2l'_T d^2 j_T} = \frac{\alpha_{\rm em}^2}{s Q^4} A_{r}
L_{\mu\nu}^r(l,\lambda_e, l^\prime) W_r^{\mu\nu}(q,p,S,j_T), \label{f:crossjperp}
\end{align}
where $\eta$ is the rapidity of the scattered lepton. We here have used $d\eta= dl'_z /E_{l'}$.

\begin{figure}
  \centering
  \includegraphics[width=0.8\linewidth]{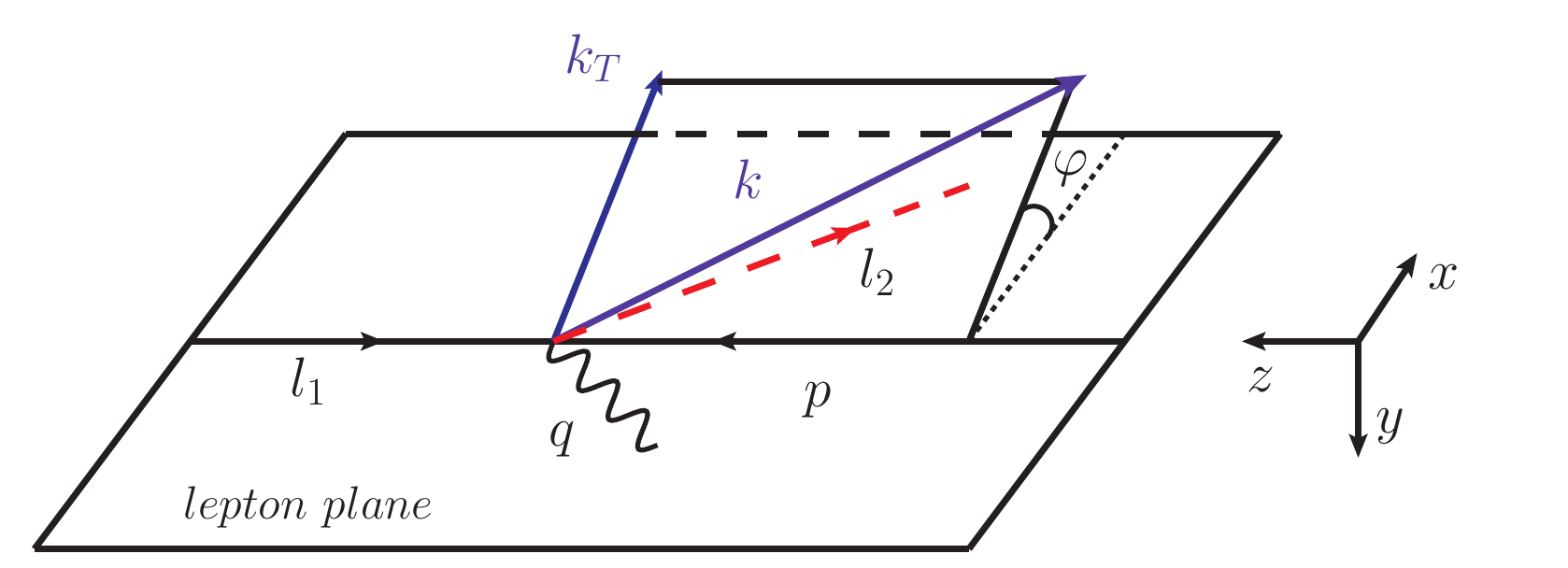}
  \caption{Illustration of the SIDIS process of the jets productions in the $eN$ collinear frame.}\label{fig:gammal}
\end{figure}

In the $eN$ collinear frame, we use the lightcone unit vectors ($\bar t, t$) to express the momenta of these particles. While in the $\gamma^*N$ collinear frame,  vectors ($\bar n, n$) are used to express those momenta. To distinguish the difference, we present the relationships in appendix \ref{sec:appendixv}.
If the target particle travels in the $+z$ direction while the incoming lepton travels in  the $-z$ direction, see Fig. \ref{fig:gammal}. We define  $\bar t^\mu=\frac{1}{\sqrt{2}}(1, 0, 0, 1), t^\mu=\frac{1}{\sqrt{2}}(1, 0, 0, -1)$. They satisfy $\bar t^2=t^2=0$, $\bar t\cdot t=1$. In lightcone coordinates, $\bar t^\mu=(1, 0, \vec 0_T), t^\mu=(0,  1, \vec 0_T)$. Therefore,
\begin{align}
  & p^\mu=p^+ \bar t^\mu + p^- t^\mu, \label{f:pmudef}\\
  & l^\mu =l^+ \bar t^\mu + l^- t^\mu, \label{f:lmudef}
\end{align}
where $p^\pm=\frac{1}{\sqrt{2}}(p_0\pm p_3)$. $l^+$ and $l^-$ are defined similarly. Up to $\mathcal{O}(1/Q^2)$, $p^\mu=p^+ \bar t^\mu,  l^\mu = l^- t^\mu$. In other words, the lightcone  vector in this frame can be defined as $\bar t^\mu=p^\mu/ p^+, t^\mu = l^\mu /l^-$.
According to the definition, we parametrize these momenta as
\begin{align}
  & p^\mu=\left(p^+, 0, \vec 0_T \right), \label{f:pmu}\\
  & l^\mu =\left(0, \frac{Q^2}{2xyp^+}, \vec 0_T \right), \label{f:lmu} \\
  & q^\mu=\left(-xyp^+, \frac{Q^2}{2xp^+}, -Q\sqrt{1-y}, 0\right), \label{f:qmu}
\end{align}
 where $x=Q^2/2p\cdot q$, $y=p\cdot q/p\cdot l$. Since the transverse momentum $\vec j_T$ equals to the intrinsic transverse momentum $\vec k_T$ of a quark in the nucleon in the  $eN$ collinear frame, we can define
\begin{align}
 j_T^\mu= k_T^\mu = |k_T|(0,0, \cos\varphi, \sin\varphi). \label{f:kmu}
\end{align}
We also parametrize the transverse polarization vector as,
\begin{align}
& S_T^\mu = |S_T| \left( 0,0, \cos\varphi_S, \sin\varphi_S \right). \label{f:Smu}
\end{align}
Here we note that the $transverse~component$ is defined with respect to the $z$-direction determined by momenta of the incoming lepton and the target hadron.

\section{The hadronic tensor} \label{sec:hadront}

In order to calculate the differential cross section, Eq. (\ref{f:crossjperp}), we need to obtain the hadronic tensor. To be explicit, we divide the hadronic tensor into a leading twist part and a twist-3 part.  Leading twist hadronic tensor had been obtained in Ref. \cite{Yang:2023vyv}, we repeat here explicitly.

\subsection{The leading twist hadronic tensor}

Equations (\ref{f:wgg})-(\ref{f:gzz}) show respectively the operator definitions of hadronic tensors for the EM, interference and weak interaction. One can choose any one of them for example for illustration. Generally, we choose $W_{ZZ}^{\mu\nu}$.  After simple algebraic calculation, the hadronic tensor  can be written as,
\begin{align}
  W^{\mu\nu}=\frac{ 2E_{k'}}{2p\cdot q}{\rm{Tr}} \left[\hat \Phi^{(0)}(x, k_T) \hat H^{\mu\nu}_{ZZ}(q,k)  \right] (2\pi)^3 \delta(q_z+k_z-k_z^\prime).
\end{align}
Hereafter, we neglect the subscript $ZZ$ of  $W^{\mu\nu}$ for simplicity. Integrating over $k_z$, see Eq. (\ref{f:hadronzz}), we have the $k_T$-dependent hadronic tensor. Changing the variable $k_T$ into $j_T$ gives $W^{\mu\nu}(q,p,S,j_T)$,
\begin{align}
   \tilde W^{\mu\nu}=\frac{1}{2p\cdot q} {\rm{Tr}} \left[\hat \Phi^{(0)}(x, k_T) \hat H^{\mu\nu}_{ZZ}(q,k) \right], \label{f:hadronlead}
\end{align}
where the quark-quark correlator is
\begin{align}
   \hat \Phi^{(0)}(x,k_T)&= \int \frac{p^+ dy^- d^2 \vec y_T}{(2\pi)^3} e^{ixp^+ y^- - i\vec k_T \vec y_T}\nonumber \\
  &\times\langle p,S| \bar \psi(0) {\cal L}(0,y) \psi(y)| p,S \rangle. \label{f:correlator}
\end{align}
The gauge link has been inserted into the quark-quark correlator Eq. (\ref{f:correlator}) to keep the gauge invariance.
In the jet-production SIDIS process, where the fragmentation is not considered, only the chiral even PDFs are involved. Since there is no spin flip, we only need to consider the $\gamma^\alpha$- and the $\gamma^\alpha\gamma^5$-terms in the decomposition of these correlators.
\begin{align}
& \hat \Phi^{(0)}= \frac{1}{2}\left[\gamma^\alpha \Phi^{(0)}_\alpha + \gamma^\alpha\gamma_5 \tilde\Phi^{(0)}_\alpha \right], \label{f:hatphi0}\\
& \hat \Phi_\rho^{(1)}= \frac{1}{2}\left[\gamma^\alpha \Phi_{\rho\alpha}^{(1)} + \gamma^\alpha\gamma_5 \tilde\Phi_{\rho\alpha}^{(1)} \right],
\end{align}
where $\hat \Phi_\rho^{(1)}$ is the quark-gluon-quark correlator which will be introduced in the next part. The TMD PDFs are defined through the decomposition of the correlators or the coefficient functions,
\begin{align}
  \Phi^{(0)}_\alpha &=p^+ \bar t_\alpha\Bigl(f_1-\frac{k_T \cdot \tilde S_T}{M}f^\perp_{1T}  \Bigr) +k_{T\alpha} f^\perp  \nonumber\\
  &- M\tilde S_{T\alpha}f_T - \lambda_h \tilde k_{T\alpha} f_L^\perp  -\frac{k_{T\langle\alpha}k_{T\beta\rangle}}{M}  \tilde S_T^\beta f_T^\perp  ,
\label{eq:Xi0Peven}\\
  \tilde\Phi^{(0)}_\alpha &=p^+\bar t_\alpha\Bigl(-\lambda_hg_{1L}+\frac{k_T\cdot S_T}{M}g^\perp_{1T}\Bigr)-\tilde k_{T\alpha}  g^\perp\nonumber\\
  &- M S_{T\alpha}g_T -\lambda_h k_{T\alpha} g_L^\perp + \frac{k_{T\langle\alpha}k_{T\beta\rangle}}{M}  S_T^\beta g_T^\perp .
\label{eq:Xi0Podd}
\end{align}
Here $\tilde A^\alpha =\varepsilon_T^{\alpha A}=\varepsilon_T^{\alpha \beta}A_{T\beta}$, $A$ can be $k_T$ or $S_T$. $k_{T\langle\alpha}k_{T\beta\rangle}= k_{T\alpha}k_{T\beta}-g_{T\alpha\beta}k_T^2/2$.
 For the antiquark distribution functions defined via the antiquark correlator $\overline{\Phi}(x,k_T) $~\cite{Mulders:1995dh,Mulders0}, we have relations $\overline{ \Phi}^{[\gamma]} =+ \Phi^{[\gamma]} $ and $\overline{ \Phi}^{[\gamma\gamma^5]} = -\Phi^{[\gamma\gamma^5]} $, where $\Phi^{[\gamma]}$ and $\Phi^{[\gamma\gamma^5]}$ denote respectively distribution functions  given in  Eq. (\ref{eq:Xi0Peven}) and (\ref{eq:Xi0Podd}).

The hard part in Eq. (\ref{f:hadronlead}) is abbreviated as
\begin{align}
  \hat H^{\mu\nu}_{ZZ}(q,k) =\Gamma^{\mu,q}(\slashed q+\slashed k) \Gamma^{\nu,q}. \label{f:hardamp}
\end{align}
In the $eN$ collinear frame we have  the following relationships
\begin{align}
 k^- << k_T << q_T \sim q^-,
\end{align}
and $q^+ + k^+=(1-y)xp^+$. Neglecting the small components of $k$, we have
\begin{align}
  (\slashed q+\slashed k) =(1-y)xp^+ \slashed{\bar t} + q^- \slashed t +\slashed q_T. \label{f:harden}
\end{align}
In addition to the minus component of $q$, we see, the transverse and the plus components also contribute in the $eN$ collinear frame. This is important to the requirement of the current conservation of the hadronic tensor. 


According to the above discussion, the hadronic tensor is calculated as
\begin{align}
 W_{t2}^{\mu\nu} =& -\left( c_1^q \tilde g_{T}^{\mu\nu} + ic_3^q \tilde \varepsilon_{T}^{\mu\nu} \right) \Bigl(f_1-\frac{k_T \cdot \tilde S_T}{M}f^\perp_{1T} \Bigr) \nonumber\\
& - \left(c_3^q \tilde g_{T}^{\mu\nu} + i c_1^q \tilde\varepsilon_{T}^{\mu\nu} \right) \Bigl(-\lambda_hg_{1L}+\frac{k_T\cdot S_T}{M}g^\perp_{1T} \Bigr), \label{f:wt2munu}
\end{align}
where subscript $t2$ denotes leading twist. Dimensionless tensors are defined as
\begin{align}
  & \tilde g_{T}^{\mu\nu} =g_{T}^{\mu\nu} -\frac{\vec q_T^2}{(q^-)^2}\bar t^\mu \bar t^\nu -\frac{1}{q^-}q_T^{\{\mu} \bar t^{\nu\}}, \label{f:gperp}\\
  & \tilde \varepsilon_{T}^{\mu\nu} = \varepsilon_{T}^{\mu\nu} +\frac{1}{q^-}\varepsilon^{\mu\nu \bar t q_T}. \label{f:eperp}
\end{align}
It is easily to check that $q_\mu \tilde{g}_T^{\mu\nu}=q_\nu \tilde{g}_T^{\mu\nu}=0$ and $q_\mu \tilde{\varepsilon}_T^{\mu\nu}=q_\nu \tilde{\varepsilon}_T^{\mu\nu}=0$. This relationships imply that the current conservation of the hadronic tensor is satisfied.

\subsection{The twist-3 hadronic tensor}

The twist-3 hadronic tensor has two parts of contributions, one from the quark-quark correlator given in Eq. (\ref{f:correlator}), the other from the quark-gluon-quark correlator,
\begin{align}
  \hat{\Phi}_{\rho}^{(1)}\left(x, k_{T}\right) &= \int \frac{p^{+} d y^{-} d^{2} y_{T}}{(2 \pi)^{3}} e^{i x p^+ y^- - i \vec{k}_T \cdot \vec{y}_T} \nonumber\\
   &\times \langle p,S| \bar{\psi}(0) D_{T \rho}(0) {\cal L}(0, y) \psi(y)| p,S\rangle, \label{f:Phi1}
\end{align}
where $D_\rho(y) = -i\partial_\rho + g A_\rho(y)$ is the covariant derivative.  Similar to Eqs. (\ref{eq:Xi1Peven}) and (\ref{eq:Xi1Podd}), we can decompose the quark-gluon-quark correlator as we did for the quark-quark correlator,
\begin{align}
  \Phi^{(1)}_{\rho\alpha}&=p^+\bar t_\alpha\Bigl[k_{T\rho} f^\perp_d- M\tilde S_{T\rho}f_{dT} \nonumber\\
  & -\lambda_h \tilde k_{T\rho} f_{dL}^\perp -\frac{k_{T\langle\rho}k_{T\beta\rangle}}{M} \tilde S_T^\beta f_{dT}^\perp \Bigr], \label{eq:Xi1Peven} \\
  \tilde \Phi^{(1)}_{\rho\alpha}&=ip^+\bar t_\alpha\Bigl[\tilde k_{T\rho}g^\perp_d+ MS_{T\rho}g_{dT}\nonumber\\
  & +\lambda_h k_{T\rho} g_{dL}^\perp  - \frac{k_{T\langle\rho}k_{T\beta\rangle}}{M} S_T^\beta g_{dT}^\perp  \Bigr],
\label{eq:Xi1Podd}
\end{align}
where a subscript $d$ is used to denote TMD PDFs defined via the quark-gluon-quark correlator or coefficient functions.

For the sake of simplicity, we show the complete twist-3 hadronic tensor here and give the calculation procedure in the appendix \ref{sec:appendixa}.
Then the complete twist-3 hadronic tensor which satisfies the current conservation is written as
\begin{align}
  W^{\mu\nu}_{t3} &=\frac{ f^\perp}{p\cdot q}h^{\mu\nu}_1 + \frac{ \lambda_h f^\perp_L}{p\cdot q}h^{\mu\nu}_2+\frac{g^\perp}{p\cdot q}h^{\mu\nu}_3 + \frac{\lambda g^\perp_L}{p\cdot q} h^{\mu\nu}_4 \nonumber \\
  &+ \frac{Mf_T}{p\cdot q}h_5^{\mu\nu} + \frac{f^\perp_T}{p\cdot q}h_6^{\mu\nu} + \frac{Mg_T}{p\cdot q}h_7^{\mu\nu} + \frac{g^\perp_T}{p\cdot q}h_8^{\mu\nu},  \label{f:wt3}
\end{align}
where $h^{\mu\nu}_{1-8}$ are defined as
\begin{widetext}
\begin{align}
   h_1^{\mu\nu}=& +c_1^q \bigg[ k_T^{\{\mu} t^{\nu\}}q^- + k_T^{\{\mu} \bar t^{\nu\}}(2-y)xp^+ + k_T^{\{\mu} q_T^{\nu\}}- \bigg( g^{\mu\nu}+\bar t^\mu \bar t^\nu \frac{2x p^+}{q^-} \bigg) k_T\cdot q_T \bigg]  \nonumber\\
  & + ic_3^q \bigg[ \tilde k_T^{[\mu} t^{\nu]}q^- -\tilde k_T^{[\mu} \bar t^{\nu]}q^+ - \bar t^{[\mu} t^{\nu]}\varepsilon_T^{qk}\bigg],  \label{f:hmunu1}\\
  h_2^{\mu\nu}=&-c_1^q \bigg[\tilde k_T^{\{\mu} t^{\nu\}}q^- + \tilde k_T^{\{\mu} \bar t^{\nu\}}(2-y)xp^+ + \tilde k_T^{\{\mu} q_T^{\nu\}}- \bigg( g^{\mu\nu}+\bar t^\mu \bar t^\nu \frac{2x p^+}{q^-} \bigg) \varepsilon_T^{q k}\bigg]  \nonumber\\
  & + ic_3^q \bigg[ k_T^{[\mu} t^{\nu]}q^- - k_T^{[\mu} \bar t^{\nu]}q^+ - \bar t^{[\mu} t^{\nu]}k_T\cdot q_T \bigg],  \label{f:hmunu2}\\
  h_3^{\mu\nu}=&-c_3^q \bigg[\tilde k_T^{\{\mu} t^{\nu\}}q^- + \tilde k_T^{\{\mu} \bar t^{\nu\}}(2-y)xp^+ + \tilde k_T^{\{\mu} q_T^{\nu\}}- \bigg( g^{\mu\nu}+\bar t^\mu \bar t^\nu \frac{2x p^+}{q^-} \bigg) \varepsilon_T^{q k}\bigg]  \nonumber\\
   & + ic_1^q \bigg[ k_T^{[\mu} t^{\nu]}q^- - k_T^{[\mu} \bar t^{\nu]}q^+ - \bar t^{[\mu} t^{\nu]}k_T\cdot q_T \bigg],  \label{f:hmunu3}\\
    h_4^{\mu\nu}=&-c_3^q \bigg[ k_T^{\{\mu} t^{\nu\}}q^- + k_T^{\{\mu} \bar t^{\nu\}}(2-y)xp^+ + k_T^{\{\mu} q_T^{\nu\}}- \bigg( g^{\mu\nu}+\bar t^\mu \bar t^\nu \frac{2x p^+}{q^-} \bigg) k_T\cdot q_T \bigg]  \nonumber\\
   & - ic_1^q \bigg[ \tilde k_T^{[\mu} t^{\nu]}q^- -\tilde k_T^{[\mu} \bar t^{\nu]}q^+ - \bar t^{[\mu} t^{\nu]}\varepsilon_T^{qk}\bigg],  \label{f:hmunu4}\\
    h_5^{\mu\nu}=&+\Bigg\{-c_1^q \bigg[\tilde S_T^{\{\mu} t^{\nu\}}q^- + \tilde S_T^{\{\mu} \bar t^{\nu\}}(2-y)xp^+ + \tilde S_T^{\{\mu} q_T^{\nu\}}- \bigg( g^{\mu\nu}+\bar t^\mu \bar t^\nu \frac{2x p^+}{q^-} \bigg) \varepsilon_T^{  q S}\bigg]  \nonumber\\
  &\quad + ic_3^q \bigg[ S_T^{[\mu} t^{\nu]}q^- - S_T^{[\mu} \bar t^{\nu]}q^+ - \bar t^{[\mu} t^{\nu]}S_T\cdot q_T \bigg]\Bigg\},  \label{f:hmunu5}\\
   h_6^{\mu\nu}=& -\Bigg\{+c_1^q \bigg[ k_T^{\{\mu} t^{\nu\}}q^- + k_T^{\{\mu} \bar t^{\nu\}}(2-y)xp^+ + k_T^{\{\mu} q_T^{\nu\}}- \bigg( g^{\mu\nu}+\bar t^\mu \bar t^\nu \frac{2x p^+}{q^-} \bigg) k_T\cdot q_T \bigg]  \nonumber\\
  &\quad + ic_3^q \bigg[ \tilde k_T^{[\mu} t^{\nu]}q^- -\tilde k_T^{[\mu} \bar t^{\nu]}q^+ - \bar t^{[\mu} t^{\nu]}\varepsilon_T^{qk}\bigg] \Bigg\}\frac{\varepsilon_T^{kS}}{M} \\
  &-\Bigg\{-c_1^q \bigg[\tilde S_T^{\{\mu} t^{\nu\}}q^- + \tilde S_T^{\{\mu} \bar t^{\nu\}}(2-y)xp^+ + \tilde S_T^{\{\mu} q_T^{\nu\}}- \bigg( g^{\mu\nu}+\bar t^\mu \bar t^\nu \frac{2x p^+}{q^-} \bigg) \varepsilon_T^{  q S}\bigg]  \nonumber\\
  &\quad + ic_3^q \bigg[ S_T^{[\mu} t^{\nu]}q^- - S_T^{[\mu} \bar t^{\nu]}q^+ - \bar t^{[\mu} t^{\nu]}S_T\cdot q_T \bigg]\Bigg\} \frac{k_T^2}{2M},   \label{f:hmunu6}\\
    h_7^{\mu\nu}=&+\Bigg\{-c_3^q \bigg[ S_T^{\{\mu} t^{\nu\}}q^- +  S_T^{\{\mu} \bar t^{\nu\}}(2-y)xp^+ +  S_T^{\{\mu} q_T^{\nu\}}- \bigg( g^{\mu\nu}+\bar t^\mu \bar t^\nu \frac{2x p^+}{q^-} \bigg)S_T\cdot q_T\bigg]  \nonumber\\
  & \quad- ic_1^q \bigg[ \tilde S_T^{[\mu} t^{\nu]}q^- -\tilde S_T^{[\mu} \bar t^{\nu]}q^+ - \bar t^{[\mu} t^{\nu]}\varepsilon_T^{  q S} \bigg]\Bigg\},  \label{f:hmunu7}\\
   h_8^{\mu\nu}=& +\Bigg\{+c_3^q \bigg[ k_T^{\{\mu} t^{\nu\}}q^- + k_T^{\{\mu} \bar t^{\nu\}}(2-y)xp^+ + k_T^{\{\mu} q_T^{\nu\}}- \bigg( g^{\mu\nu}+\bar t^\mu \bar t^\nu \frac{2x p^+}{q^-} \bigg) k_T\cdot q_T \bigg]  \nonumber\\
  &\quad + ic_1^q \bigg[ \tilde k_T^{[\mu} t^{\nu]}q^- -\tilde k_T^{[\mu} \bar t^{\nu]}q^+ - \bar t^{[\mu} t^{\nu]}\varepsilon_T^{qk}\bigg] \Bigg\}\frac{k_T\cdot S_T}{M} \\
  &-\Bigg\{+c_3^q \bigg[ S_T^{\{\mu} t^{\nu\}}q^- +  S_T^{\{\mu} \bar t^{\nu\}}(2-y)xp^+ +  S_T^{\{\mu} q_T^{\nu\}}- \bigg( g^{\mu\nu}+\bar t^\mu \bar t^\nu \frac{2x p^+}{q^-} \bigg)S_T\cdot q_T\bigg]  \nonumber\\
  & \quad+ic_1^q \bigg[ \tilde S_T^{[\mu} t^{\nu]}q^- -\tilde S_T^{[\mu} \bar t^{\nu]}q^+ - \bar t^{[\mu} t^{\nu]}\varepsilon_T^{  q S} \bigg]\Bigg\}\frac{k_T^2}{2M}.   \label{f:hmunu8}
\end{align}
\end{widetext}
We see clearly that the full twist-3 hadronic tensor satisfies current conservation, $q_\mu \tilde W^{\mu\nu}_{t3} = q_\nu \tilde W^{\mu\nu}_{t3} = 0$. Although, the $h$-tensors are relative complicated, they have the similar forms which would lead to the simple expression of the differential cross section.

\section{The Results} \label{sec:results}

\subsection{The differential cross section}

The differential cross section can be obtained by using the contraction of the leptonic tensor and hadronic tensor. With variables shown in Eqs. (\ref{f:pmu})-(\ref{f:Smu}), we  use  Eqs. (\ref{f:leptonzz}) and (\ref{f:gperp}), (\ref{f:eperp}) and obtain
\begin{align}
  & L_{\mu\nu}^{ZZ} \left(c_1^q \tilde g_T^{\mu\nu} + ic_3^q \varepsilon_T^{\mu\nu}\right) = -\frac{2Q^2}{y^2}\left[T_0^q(y)-\lambda_e \tilde T_0^q(y)\right], \\
  & L_{\mu\nu}^{ZZ} \left(c_3^q \tilde g_T^{\mu\nu} + ic_1^q \varepsilon_T^{\mu\nu}\right) = -\frac{2Q^2}{y^2}\left[\tilde T_1^q(y)-\lambda_e T_1^q(y)\right],
\end{align}
where $T$-functions are defined as
\begin{align}
  & T_0^q(y) = c_1^e c_1^q A(y) + c_3^e c_3^q C(y), \nonumber\\
  & \tilde T_0^q(y) = c_3^e c_1^q A(y) + c_1^e c_3^q C(y), \nonumber\\
  & T_1^q(y) = c_3^e c_3^q A(y) + c_1^e c_1^q C(y), \nonumber\\
  & \tilde T_1^q(y) = c_1^e c_3^q A(y) + c_3^e c_1^q C(y). \label{eq:T0T1}
\end{align}
Here $A(y)=y^2-2y+2, C(y)=y(2-y)$. A simple algebraic calculation gives  the leading twist cross section of the jet-production SIDIS in the $eN$ collinear frame
\begin{align}
  d\tilde \sigma_{t2}&= \frac{\alpha_{\rm em}^2 \chi}{y Q^4}2x\Biggl\{ \left(T_0^q(y) -\lambda_e \tilde T_0^q(y) \right) f_1 \nonumber\\
   &\quad\quad - \left(\tilde T_1^q(y) -\lambda_e T_1^q(y) \right) \lambda_h g_{1L} \nonumber\\
  &+|S_T|k_{T M}\Big[\sin(\varphi-\varphi_S) \left(T_0^q(y) -\lambda_e \tilde T_0^q(y) \right) f^\perp_{1T} \nonumber \\
  &\quad  -\cos(\varphi-\varphi_S) \left(\tilde T_1^q(y) -\lambda_e T_1^q(y) \right) g^\perp_{1T}\Big] \Biggr\}, \label{f:crosslead}
\end{align}
where  $d\tilde \sigma_{t2}=d\sigma_{t2}/(d\eta d^2l'_T d^2 j_T)$,  $k_{T M}=|k_T|/M$. Subscript $t2$ denotes leading twist.

We can calculate the  twist-3 differential cross section similarly.  For the sake of simplicity, we only show  contractions of the leptonic tensor and $h^{\mu\nu}_{1,3}$ here,
\begin{align}
  & L_{\mu\nu}^{ZZ}\cdot h_1^{\mu\nu} = -\frac{2Q^3}{y^2}|k_T| \left[T_2^q(y)-\lambda_e \tilde T_2^q(y)\right]\cos\varphi, \\
  & L_{\mu\nu}^{ZZ}\cdot h_3^{\mu\nu} = -\frac{2Q^3}{y^2}|k_T| \left[\tilde T_3^q(y)-\lambda_e T_3^q(y)\right]\sin\varphi.
\end{align}
Other contractions have the same forms. The $T$-functions are defined as
\begin{align}
  & T_2^q(y) = c_1^e c_1^q B(y) + c_3^e c_3^q D(y), \nonumber\\
  & \tilde T_2^q(y) = c_3^e c_1^q B(y) + c_1^e c_3^q D(y), \nonumber\\
  & T_3^q(y) = c_3^e c_3^q B(y) + c_1^e c_1^q D(y), \nonumber\\
  & \tilde T_3^q(y) = c_1^e c_3^q B(y) + c_3^e c_1^q D(y). \label{eq:T2T3}
\end{align}
with $B(y)=2-y^2, D(y)=y^2\sqrt{1-y}$. After simple calculations, we write down the differential cross section at twist-3,
\begin{align}
 d\tilde \sigma_{t3} &= -\frac{\alpha_{\rm em}^2 \chi}{y Q^4} 4x^2 \kappa_M \Biggl\{k_{T M}  \cos\varphi\left(T_2^q(y) -\lambda_e \tilde T_2^q(y) \right) f^\perp \nonumber\\
  &\quad + k_{T M}\sin\varphi\left(\tilde T_3^q(y) -\lambda_e T_3^q(y) \right) g^\perp \nonumber\\
 & -\lambda_h \Big[  k_{T M}\cos\varphi \left(\tilde T_3^q(y) -\lambda_e T_3^q(y) \right)g^\perp_{L} \nonumber\\
  &\quad -k_{T M} \sin\varphi\left(T_2^q(y)   -\lambda_e \tilde T_2^q(y) \right) f^\perp_{L}  \Big] \nonumber \\
  + & |S_T|\Big[\sin\varphi_S \left(T_2^q(y) -\lambda_e \tilde T_2^q(y) \right)f_T \nonumber\\
  &\quad +\cos\varphi_S\left(\tilde T_3^q(y) -\lambda_e T_3^q(y) \right)  g_T \nonumber \\
  &+\sin(2\varphi-\varphi_S) \left(T_2^q(y) -\lambda_e \tilde T_2^q(y) \right)\frac{k_{T M}^2}{2}f_T^\perp \nonumber \\
 & +  \cos(2\varphi-\varphi_S)\left(\tilde T_3^q(y) -\lambda_e T_3^q(y) \right) \frac{k_{T M}^2}{2}g_T^\perp \Big]\Biggr\},
\end{align}
 where  $d\tilde \sigma_{t3}=d\sigma_{t3}/(d\eta d^2l'_T d^2 j_T)$, $\kappa_M=M/Q$ is the twist suppression factor. In the $eN$ collinear frame, $k_T=k^\prime_T+l^\prime_T$. In the $\gamma^*N$ collinear frame, $k_\perp=k^\prime_\perp$ and $d\tilde \sigma_{t3}$ is defined as $d\tilde \sigma_{t3}=d\sigma_{t3}/(dxdyd\psi d^2k_\perp^\prime)$, where $\psi$ is the azimuthal angle of $\vec l^\prime$ around $\vec l$ \cite{Yang:2022xwy}. The difference of the transverse momentum of the incident quark leading to the different expressions of the cross section at twist-3 level. To be precise, $B(y)$ and $D(y)$ are different in these two frame. For example, for the $f^\perp$ term, we have $2Q^3|k_\perp| \cos\varphi^\prime 2(2-y)\sqrt{1-y}/y^2$ in the $\gamma^* N$ collinear frame and $2Q^3|k_T| \cos\varphi (2-y^2)/y^2$ in the $eN$ collinear frame.

In this part, we only present results for the weak interaction case. For the EM contribution, it requires $c_3^{e/q} = 0$ and $c_1^{e/q} = 1$.  For the interference contribution, it requires  $c_3^{e/q}=c_A^{e/q}$ and $c_1^{e/q}=c_V^{e/q}$.
Besides, the kinematic factors are also different. To make it transparent, we can get the EM and interference cross sections by replacing the parameters in the weak interaction cross section according to Tab.~\ref{tab:replacing}.

\begin{table}
\renewcommand\arraystretch{1.5}
\begin{tabular}{c|c|c|c}
\hline
~~~~  & $A_r$   & $L^{\mu\nu}_r$   & $W^{\mu\nu}_r$  \\ \hline
~ $ZZ$~ & $\chi$ & $c_1^e,~c_3^e$ & $c_1^q,~c_3^q$  \\
$\gamma Z$  & ~$\chi\to \chi_{int}$ ~& ~~$c_1^e\to c_V^e,~c_3^e\to c_A^e$~~ &~~ $c_1^q\to c_V^q,~c_3^q\to c_A^q$~~ \\
$\gamma\gamma$ & $\chi\to e_q^2$  & $c_1^e\to 1,~c_3^e\to 0$  & $c_1^q\to 1,~c_3^q\to 0$ \\ \hline
\end{tabular}
\caption{Relations of kinematic factors and electroweak couplings between weak, EM and interference interactions.}
\label{tab:replacing}
\end{table}

\subsection{The azimuthal asymmetries}

Azimuthal symmetries are measurable quantities which are generally used to extract (TMD) PDFs. In the reaction of jets production, the soft parts are only TMD PDFs. Uncertainties from FFs vanish. Under this circumstance, jet-production SIDIS process can be good a reaction in determining the TMD PDFs.

As to azimuthal asymmetries,  we consider both the unpolarized beam ($\lambda_e=0$) and the polarized beam ($\lambda_e=\pm1$) cases. They contribute to different azimuthal asymmetries results.
The azimuthal asymmetry has a definite definition, e.g.,
\begin{align}
  \langle \sin\varphi \rangle_{U,U}=\frac{\int d\tilde\sigma \sin\varphi d\varphi}{\int d\tilde\sigma d\varphi},
\end{align}
for the unpolarized or longitudinally polarized target case, and
\begin{align}
  \langle \sin(\varphi-\varphi_S) \rangle_{U,T}=\frac{\int d\tilde\sigma \sin(\varphi-\varphi_S)d\varphi d\varphi_S}{\int d\tilde\sigma d\varphi d\varphi_S},
\end{align}
for the transversely polarized target case.
The subscripts such as ($U, T$) denote the polarizations of the lepton beam and the target, respectively.
At the leading twist, there are two polarization dependent azimuthal asymmetries which are given by (the sum over $r=ZZ$, $\gamma Z$ and $\gamma\gamma$ is implicit in the numerator and the denominator respectively)
\begin{align}
 & \langle \sin(\varphi-\varphi_S) \rangle_{U,T} = -k_{T M} \frac{ \chi T_{0 }^q(y) f^\perp_{1T}}{2\chi T_{0 }^q(y) f_1}, \\
 & \langle \cos(\varphi-\varphi_S) \rangle_{U,T} =  k_{T M} \frac{\chi \tilde T_{1 }^q(y)g^\perp_{1T}}{2\chi T_{0 }^q(y) f_1}.
\end{align}
$f^\perp_{1T}$ is the famous Sivers function\cite{Sivers:1989cc,Sivers:1990fh} which has been studied widely.  At twist-3, we have 8 azimuthal asymmetries. They are given by
\begin{align}
  & \langle \cos\varphi \rangle_{U,U} = -x\kappa_M k_{T M} \frac{ \chi T_{2 }^q(y)}{\chi T_{0 }^q(y)}\frac{f^\perp}{f_1}, \\
  & \langle \sin\varphi \rangle_{U,U} = -x\kappa_M k_{T M} \frac{\chi \tilde T_{3 }^q(y)}{\chi T_{0 }^q(y)}\frac{g^\perp}{f_1}, \\
  & \langle \cos\varphi \rangle_{U,L} = -x\kappa_M k_{T M} \frac{\chi T_{2 }^q(y) f^\perp -\lambda_h \chi \tilde T_{3 }^q(y)g^\perp_L}{\chi T_{0 }^q(y) f_1}, \\
  & \langle \sin\varphi \rangle_{U,L} =- x\kappa_M k_{T M} \frac{\chi \tilde T_{3 }^q(y) g^\perp + \lambda_h \chi T_{2 }^q(y)f^\perp_L}{\chi T_{0 }^q(y)f_1}, \\
  & \langle \cos\varphi_S \rangle_{U,T} =-  x\kappa_M\frac{ \chi \tilde T_{3 }^q(y)g_T}{\chi T_{0 }^q(y)f_1}, \\
  & \langle \sin\varphi_S \rangle_{U,T} =- x\kappa_M\frac{ \chi T_{2 }^q(y)f_T}{\chi T_{0 }^q(y)f_1}, \\
  & \langle \cos(2\varphi-\varphi_S) \rangle_{U,T} = x\kappa_M k_{T M}^2\frac{ \chi \tilde T_{3 }^q(y)g^\perp_T}{2\chi T_{0 }^q(y)f_1}, \\
  & \langle \sin(2\varphi-\varphi_S) \rangle_{U,T} =- x\kappa_M k_{T M}^2\frac{ \chi T_{2 }^q(y)f^\perp_T}{2\chi T_{0 }^q(y)f_1}.
\end{align}

For the case of the polarized electron beam, we obtain similar results as the unpolarized case.
They have one-to-one correspondence. The two kinds of asymmetries at  leading twist  are given by,
\begin{align}
 & \langle \sin(\varphi-\varphi_S) \rangle_{L,T} = \lambda_ek_{T M} \frac{ \chi \tilde T_{0 }^q(y)f^\perp_{1T}}{2\chi T_{0 }^q(y)f_1}, \\
 & \langle \cos(\varphi-\varphi_S) \rangle_{L,T} =  -\lambda_ek_{T M} \frac{ \chi T_{1 }^q(y)g^\perp_{1T}}{2\chi T_{0 }^q(y)f_1}.
\end{align}
At twist-3, we have 8 azimuthal asymmetries. They are given by
\begin{align}
  & \langle \cos\varphi \rangle_{L,U} = \lambda_e x\kappa_M k_{T M} \frac{\chi \tilde T_{2 }^q(y)f^\perp}{\chi T_{0 }^q(y)f_1}, \\
  & \langle \sin\varphi \rangle_{L,U} = \lambda_e x\kappa_M k_{T M} \frac{ \chi T_{3 }^q(y)g^\perp}{\chi T_{0 }^q(y)f_1}, \\
  & \langle \cos\varphi \rangle_{L,L} = \lambda_e x\kappa_M k_{T M} \frac{ \chi \tilde T_{2 }^q(y) f^\perp - \lambda_h \chi T_{3 }^q(y)g^\perp_L}{\chi T_{0 }^q(y)f_1}, \\
  & \langle \sin\varphi \rangle_{L,L} = \lambda_e x\kappa_M k_{T M} \frac{\chi T_{3 }^q(y) g^\perp + \chi \tilde T_{2 }^q(y)f^\perp_L}{\chi T_{0 }^q(y)f_1}, \\
  & \langle \cos\varphi_S \rangle_{L,T} = - \lambda_e x\kappa_M\frac{ \chi T_{3 }^q(y)g_T}{\chi T_{0 }^q(y)f_1}, \\
  & \langle \sin\varphi_S \rangle_{L,T} = - \lambda_e x\kappa_M\frac{ \chi \tilde T_{2 }^q(y)f_T}{\chi T_{0 }^q(y)f_1}, \\
  & \langle \cos(2\varphi-\varphi_S) \rangle_{L,T} = \lambda_e x\kappa_M k_{T M}^2\frac{ \chi T_{3 }^q(y)g^\perp_T}{2\chi T_{0 }^q(y)f_1}, \\
  & \langle \sin(2\varphi-\varphi_S) \rangle_{L,T} = -\lambda_e x\kappa_M k_{T M}^2\frac{\chi \tilde T_{2 }^q(y)f^\perp_T}{2\chi T_{0 }^q(y)f_1}.
\end{align}

In the neutral current SIDIS process, weak contributions can not be separated from the EM contribution. We have calculated the contribution from the EM interaction. Numerical estimates shows that weak contributions will reach a few percent when $Q>10$ GeV. However, the precise values depend on the fraction $x$ and $y$. Under this circumstance, weak contributions should be taken into account in measurements of these asymmetries in the SIDIS process.

\subsection{The intrinsic asymmetries}

In the $eN$ collinear frame, $\vec j_T  =\vec k_T$ if the higher order radiation of gluon is neglected, see Eq. (\ref{f:jetk}). In other words, the transverse momentum $\vec j_T$ equals to the intrinsic transverse momentum $\vec k_T$ of a quark in the nucleon. To explore the imbalance of the transverse momentum of the incident quark in a nucleon, we introduce the intrinsic asymmetry \cite{Yang:2022xwy}.
According to the definition, the transverse momentum of the incident quark (jet) is in the x-y plane. It can be decomposed as
\begin{align}
 & k_T^{x}=k_T \cos\varphi, \\
 & k_T^{y}=k_T \sin\varphi.
\end{align}
Therefore, we can define $k_T^{x} (-x)-k_T^{x} (+x)$ to quantify the difference of the transverse momentum between the negative $x$ and positive $x$ directions. The difference in the $y$-direction is defined similarly. To be explicit, we present the general definition of the intrinsic asymmetry,
\begin{align}
A^x &= \frac{\int_{\pi/2}^{3\pi/2} d\varphi~ d\tilde{\sigma}-\int_{-\pi/2}^{\pi/2} d\varphi ~d\tilde{\sigma} }{\int_{-\pi/2}^{\pi/2} d\varphi ~d\tilde{\sigma} +\int_{\pi/2}^{3\pi/2} d\varphi~ d\tilde{\sigma}}, \label{f:akx}\\
A^y &= \frac{\int_{\pi}^{2\pi} d\varphi~ d\tilde{\sigma}-\int_{0}^{\pi} d\varphi ~d\tilde{\sigma} }{\int_{0}^{\pi} d\varphi ~d\tilde{\sigma} +\int_{\pi}^{2\pi} d\varphi~ d\tilde{\sigma}}. \label{f:aky}
\end{align}
The sum of the differential cross section for EM, weak and interference terms is understood. Equations (\ref{f:akx}) and (\ref{f:aky}) lead to asymmetries in the $x$-direction and $y$-direction, respectively.

According to our definition, the twist-3 intrinsic  asymmetries are obtained as
\begin{align}
 & A_{U,U}^x =  \frac{4x\kappa_M k_{T M}}{\pi} \frac{\chi T^q_2(y)f^\perp}{\chi T^q_0(y) f_1}, \label{f:auux} \\
 & A_{U,U}^y =  \frac{4x\kappa_M k_{T M}}{\pi} \frac{\chi \tilde T^q_3(y)g^\perp}{\chi T^q_0(y) f_1}, \label{f:auuy} \\
 & A_{U,L}^x =  -\frac{4x\kappa_M k_{T M}}{\pi} \frac{\chi \tilde T^q_3(y)g_L^\perp}{\chi T^q_0(y) f_1}, \label{f:aulx} \\
 & A_{U,L}^y =  \frac{4x\kappa_M k_{T M}}{\pi} \frac{\chi T^q_2(y)f_L^\perp}{\chi T^q_0(y) f_1}, \label{f:auly} \\
 & A_{L,U}^x =- \frac{4x\kappa_M k_{T M}}{\pi} \frac{\chi \tilde T^q_2(y)f^\perp}{\chi T^q_0(y) f_1}, \label{f:alux} \\
 & A_{L,U}^y =- \frac{4x\kappa_M k_{T M}}{\pi} \frac{\chi T^q_3(y)g^\perp}{\chi T^q_0(y) f_1}, \label{f:aluy} \\
 & A_{L,L}^x = \frac{4x\kappa_M k_{T M}}{\pi} \frac{\chi T^q_3(y)g_L^\perp}{\chi T^q_0(y) f_1}, \label{f:allx} \\
 & A_{L,L}^y =- \frac{4x\kappa_M k_{T M}}{\pi} \frac{\chi \tilde T^q_2(y)f_L^\perp}{\chi T^q_0(y) f_1}. \label{f:ally}
\end{align}
We note again that only weak interaction results are shown in Eqs. (\ref{f:auux})-(\ref{f:ally}). For the complete results, EM and interference interactions should be included. Furthermore, the sum of the quark flavor in the numerators and denominators is also understood. At leading twist, the intrinsic asymmetries can also be introduced. But they do not have physical interpretations. We do not consider them here.

In order to illustrate the intrinsic asymmetries shown above, we present the numerical values of $A_{U,U}^x$ and $A_{L,U}^x$ in Figs. \ref{fig:Axuuu} and \ref{fig:Axuud}. Without proper parametrizations, our estimations are based on the Gaussian ansatz for TMD PDFs, i.e.,
\begin{align}
 & f_1(x, k_T) =\frac{1}{\pi {\Delta^\prime}^2} f_1(x) e^{-\vec k_T^2/{\Delta^\prime}^2}, \\
 & f^\perp(x, k_T) =\frac{1}{\pi \Delta^2x}f_1(x) e^{-\vec k_T^2/\Delta^2},
\end{align}
where $f_1(x)$ are taken from CT14 \cite{Schmidt:2015zda} and the faction is taken as $x=0.3$ for illustration. We have used the Wandzura-Wilczek approximation, i.e., neglecting quark-gluon-quark correlation function ($g=0$) \cite{Mulders:1995dh,Bacchetta:2006tn}, to determine $f^\perp(x, k_T)$.

In the numerical estimates, only the valence quarks are taken into account. Because other  contributions are small. For the Gaussian ansatz the widths of the unpolarized TMD PDF $f_1(x,k_T)$ are taken as ${\Delta^\prime}^2_u={\Delta^\prime}^2_d= 0.53~\rm{GeV}^2 $ \cite{Anselmino:2005nn,Signori:2013mda,Anselmino:2013lza,Cammarota:2020qcw,Bacchetta:2022awv}. However, the widths of the unpolarized TMD PDF $f^\perp(x,k_T)$ run from $0.3$ to $0.6~\rm{GeV}^2$, see Figs. \ref{fig:Axuuu} and \ref{fig:Axuud}.
Figure \ref{fig:Axuuu}  showd the results at $\Delta_{u}^2=0.5~\rm{GeV}^2 $ while Fig. \ref{fig:Axuud}  showd the results at $\Delta_{d}^2=0.5~\rm{GeV}^2 $. In both figures, we choose $y=0.5$.

According to the numerical estimates, we find that  asymmetry $A_{L,U}^x$ is two or three orders of magnitude smaller than $A_{U,U}^x$. Because $A_{L,U}^x$ is a parity violating effect or an effect of the weak interaction. It should be the same order of magnitude as parity violation in standard model. In addition, asymmetry $A_{U,U}^x$  decreases with respect to the energy, while $A_{L,U}^x$ increases with the energy. Furthermore,  we find the intrinsic asymmetry is more sensitive to $\Delta_{u}^2$ than $\Delta_{d}^2$. We attribute it to the fact that $f^\perp(x,k_T)$  for $u$ quark is larger than that for the $d$ quark in the Gaussian ansatz and  the small variation of $\Delta_{u}^2$ will be magnified to the intrinsic asymmetry due to the larger distribution function.

\begin{widetext}

\begin{figure}
\centering
\includegraphics[width= 0.4\linewidth]{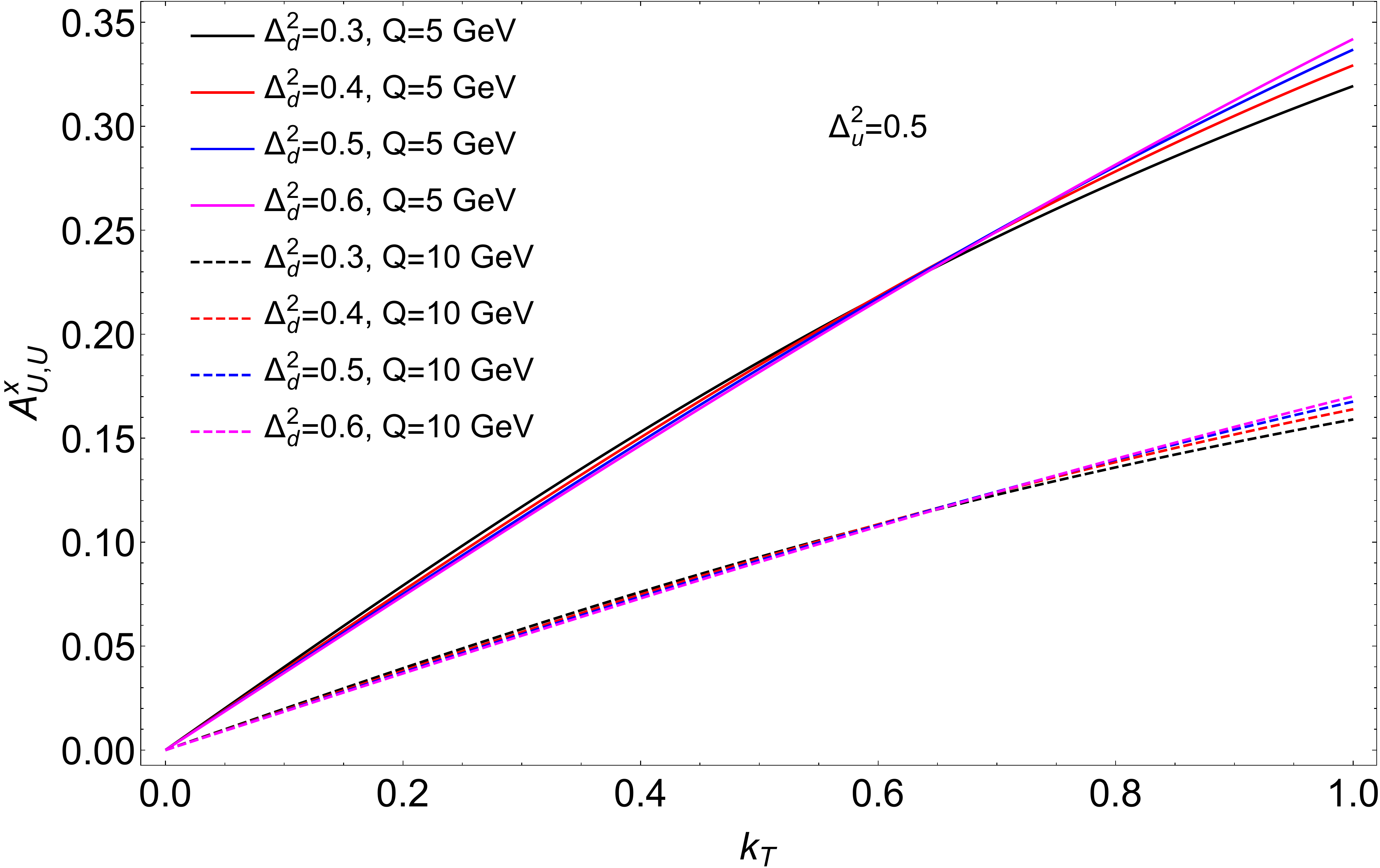}
\quad \quad \quad
\includegraphics[width= 0.425\linewidth]{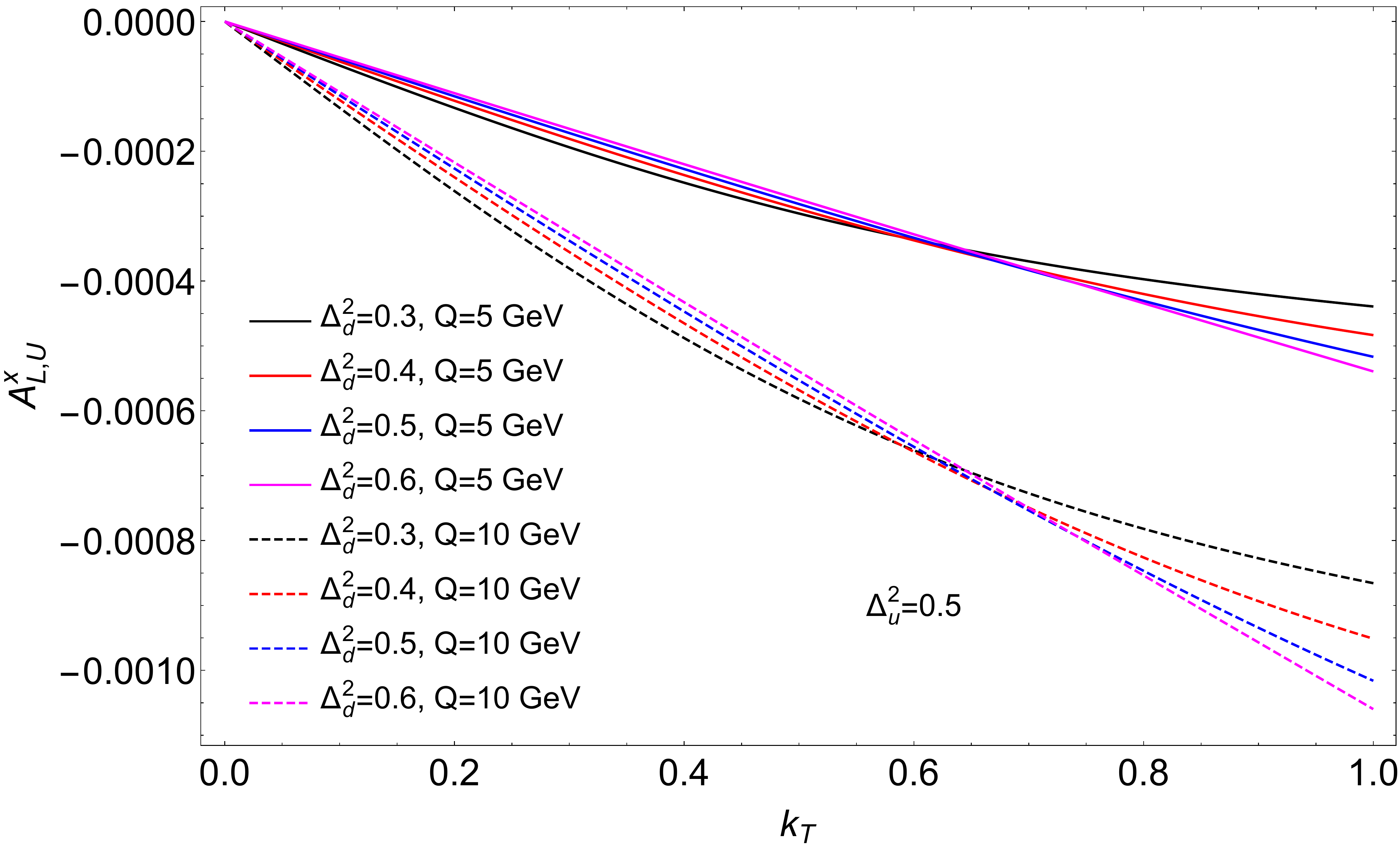}
\caption{The intrinsic asymmetry $A_{U,U}^x$ with respect to $k_T$. The solid lines show the asymmetry at 5~GeV while the dashed lines show the asymmetry at $Q=$10~GeV. Here ${\Delta^\prime}^2_u={\Delta^\prime}^2_d= 0.53~\rm{GeV}^2 $~ and $\Delta_{u}^2= 0.5~\rm{GeV}^2 $ while $\Delta_{d}^2$ runs from $0.3$ to $0.6 ~\rm{GeV}^2 $.}
\label{fig:Axuuu}
\end{figure}

\begin{figure}
\centering
\includegraphics[width= 0.4\linewidth]{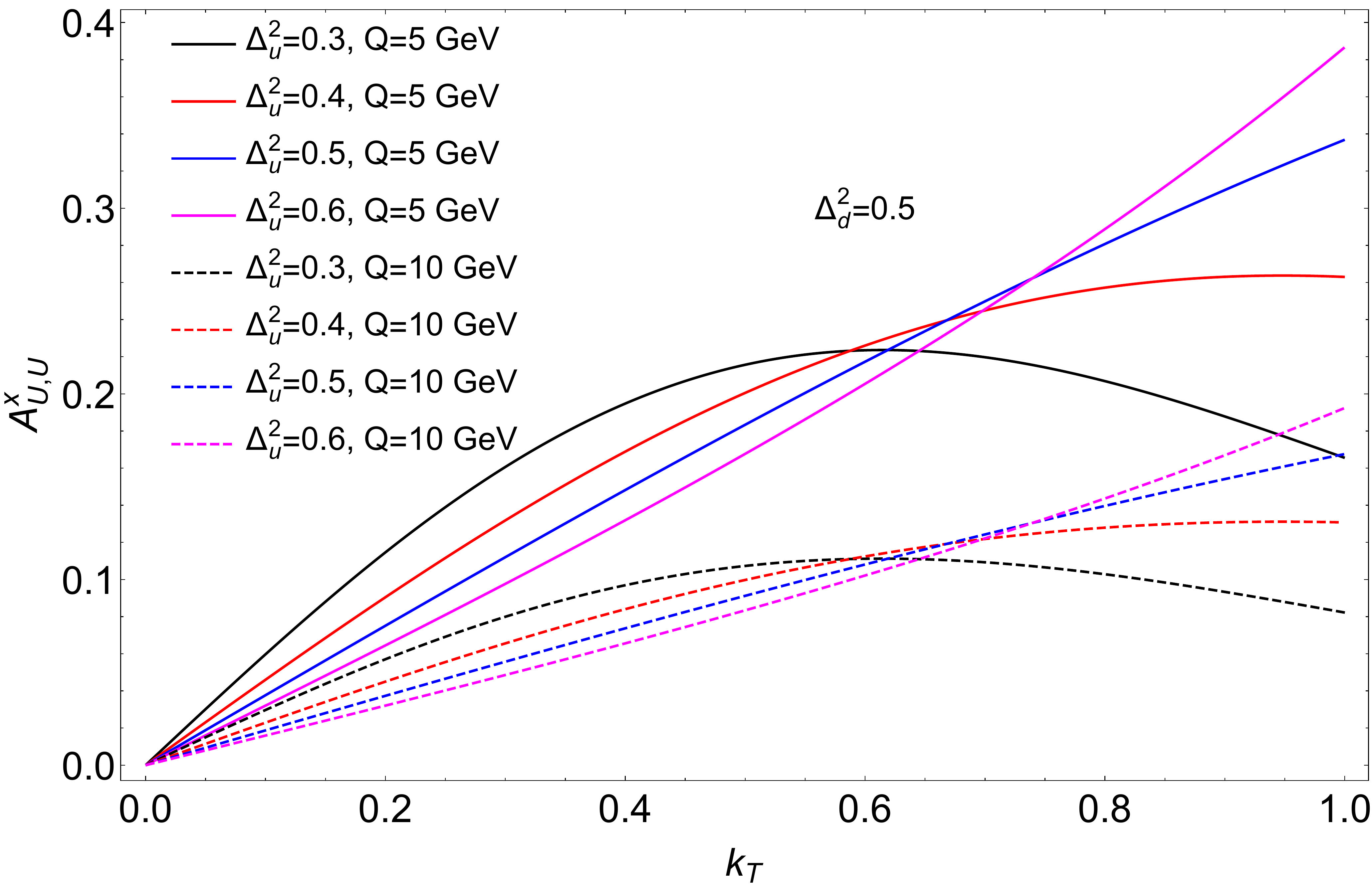}
\quad \quad \quad
\includegraphics[width= 0.43\linewidth]{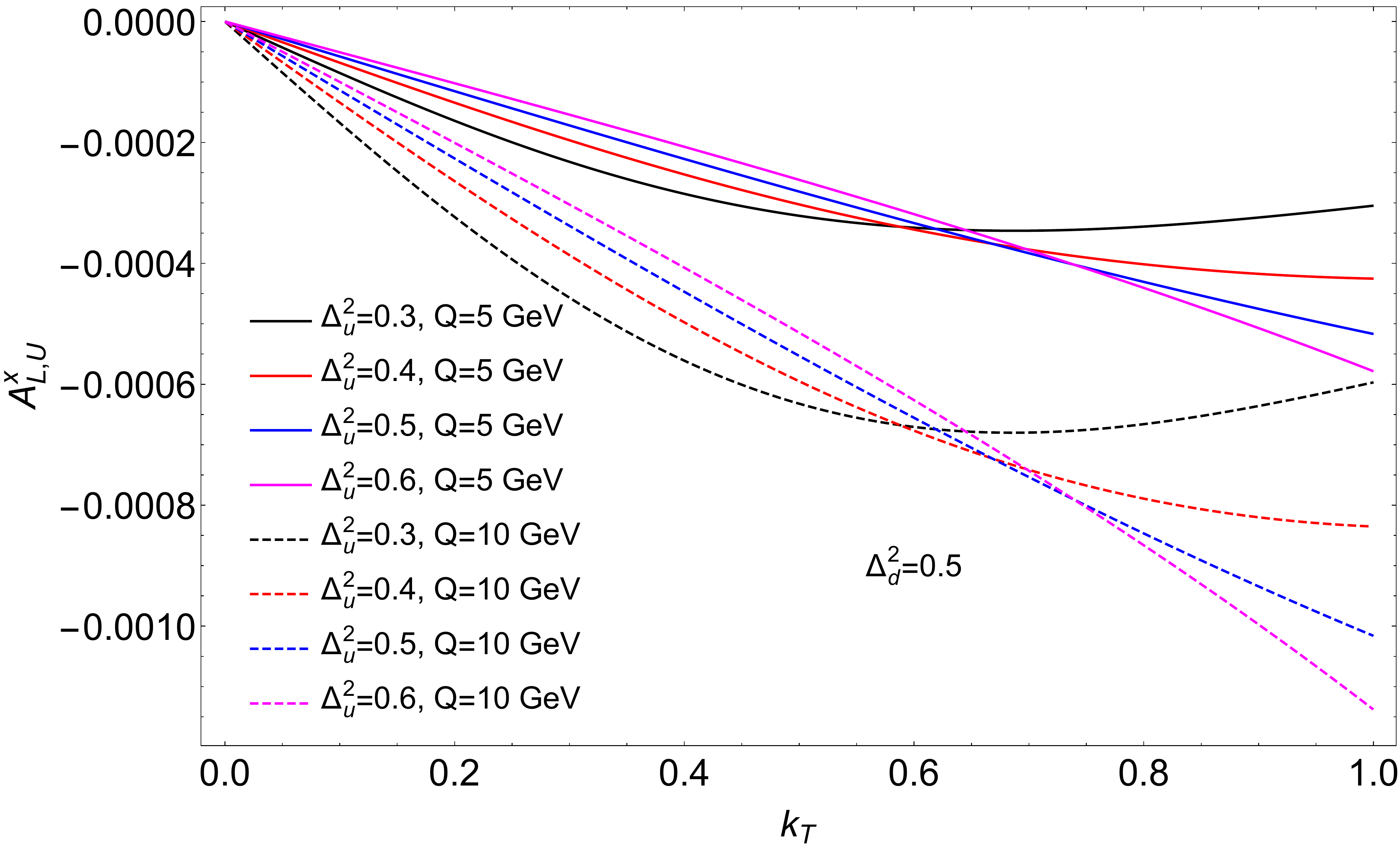}
\caption{The intrinsic asymmetry $A_{L,U}^x$ with respect to $k_T$. The solid lines show the asymmetry at 5~GeV while the dashed lines show the asymmetry at $Q=$10~GeV. Here ${\Delta^\prime}^2_u={\Delta^\prime}^2_d= 0.53~\rm{GeV}^2 $~ and $\Delta_{u}^2= 0.5~\rm{GeV}^2 $ while $\Delta_{d}^2$ runs from $0.3$ to $0.6 ~\rm{GeV}^2 $.}
\label{fig:Axuud}
\end{figure}

\end{widetext}

\section{Summary} \label{sec:summary}

In this paper, we consider the neutral current jet-production SIDIS process and calculate the differential cross section of this process at tree level twist-3 in the $eN$ collinear frame.  In this frame, the virtual-photon gains the transverse momentum component $q_T$ and the current conservation law becomes complicated.
Our calculation includes the EM, weak and inference interactions. The initial electron is assumed to be polarized and then scattered off by a target particle with spin-1/2.

After obtaining the differential cross section, we calculate azimuthal asymmetries and intrinsic asymmetries. They provide more measurable quantities for extracting (TMD) PDFs.  Two leading twist and eight twist-3 azimuthal asymmetries are obtained for the  case of the unpolarized electron beam. Similar results for the case of the polarized electron beam are also obtained. Intrinsic asymmetries indicate  the  imbalance of the transverse momentum of the incident quark in a nucleon. From the numerical estimates, we find that asymmetry $A_{L,U}^x$ is two or three orders of magnitude smaller than $A_{U,U}^x$ because of the parity violating effect. In addition, asymmetry $A_{U,U}^x$  decreases with respect to the energy, while $A_{L,U}^x$ increases with the energy. Furthermore,  the intrinsic asymmetry is more sensitive to $\Delta_{u}^2$ than $\Delta_{d}^2$.

\section*{Acknowledgements}
The author thanks X. H. Yang very much for his kind help. This work was supported by the Natural Science Foundation of Shandong Province (Grant No. ZR2021QA015).

\begin{appendix}

\section{Relationships of the lightcone vectors} \label{sec:appendixv}

\begin{figure}
  \centering
  \includegraphics[width=0.8\linewidth]{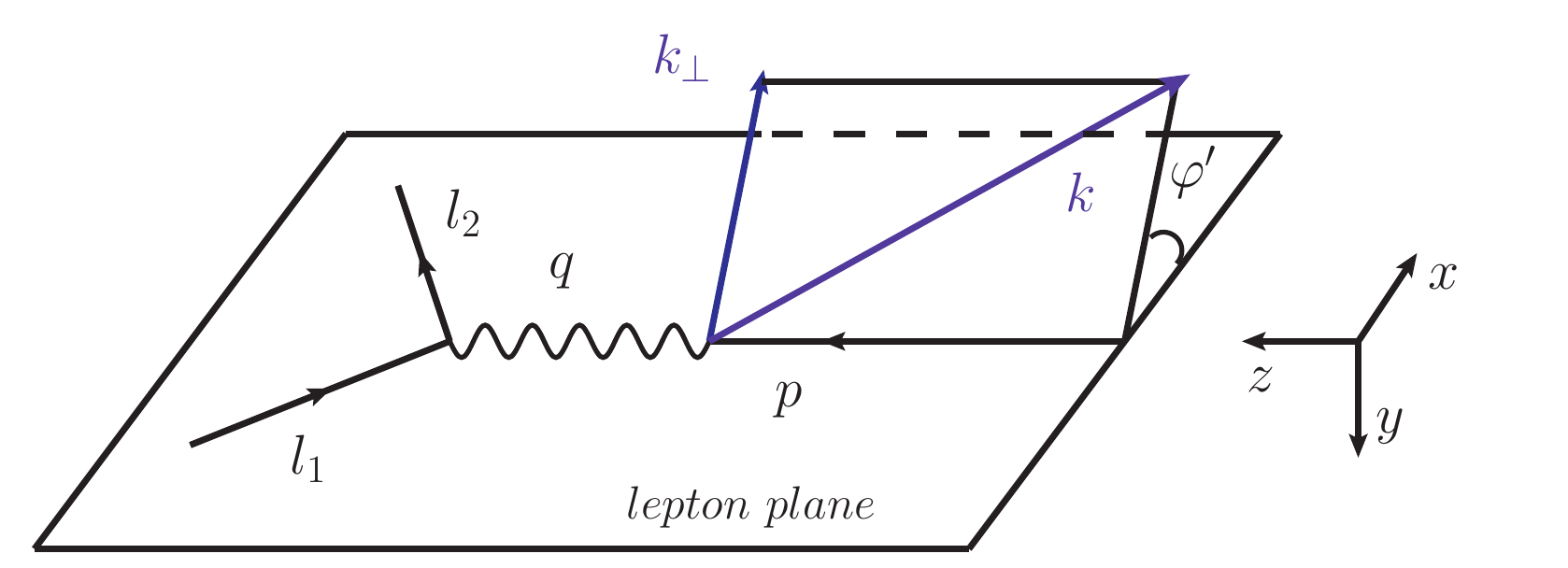}
  \caption{Illustration of the SIDIS process of the jets productions in the $\gamma^*N$ collinear frame.}\label{fig:gammap}
\end{figure}

In the $\gamma^* N$ collinear frame, see Fig. \ref{fig:gammap}, the target particle travels in the $+z$ direction.
We define  $\bar n^\mu=\frac{1}{\sqrt{2}}(1, 0, 0, 1), n^\mu=\frac{1}{\sqrt{2}}(1, 0, 0, -1)$. They satisfy $\bar n^2=n^2=0$, $\bar n\cdot n=1$. In lightcone coordinates, $\bar n^\mu=(1, 0, \vec 0_T), n^\mu=(0,  1, \vec 0_T)$. Therefore,
\begin{align}
  & p^\mu=\tilde p^+ \bar n^\mu + \tilde p^- n^\mu, \label{f:ptildemudef}\\
  & q^\mu =q^+ \bar n^\mu + q^- n^\mu, \label{f:qmudef}
\end{align}
Here the plus component of $p^\mu$ is written as $\tilde p$ to distinguish that in the $eN$ collinear frame.  Up to $\mathcal{O}(1/Q^2)$, only the large plus component of $p^\mu$ survives. Lightcone vectors can be defined as $\bar n^\mu=p^\mu/\tilde p^+, n^\mu=(q^\mu+xp^\mu)/q^-$.
Under this condition, we parametrize momenta of these particles as \cite{Chen:2020ugq}
\begin{align}
& p^\mu = \left(\tilde p^+,0,\vec 0_\perp \right), \nonumber\\
& q^\mu = \left( -x\tilde p^+, \frac{Q^2}{2x\tilde p^+}, \vec 0_\perp \right),\nonumber\\
& l^\mu = \left( \frac{1-y}{y}x\tilde p^+, \frac{Q^2}{2xy\tilde p^+}, \frac{Q\sqrt{1-y}}{y},0 \right),\nonumber\\
& l^{\prime \mu}=\left( \frac{1}{y}x\tilde p^+, \frac{(1-y)Q^2}{2xy\tilde p^+}, \frac{Q\sqrt{1-y}}{y},0 \right).
\end{align}

The relationships between $\bar n, n$ and $\bar t, t$ are
\begin{align}
 & \bar t^\mu = \frac{(2-y)}{y}\frac{\tilde p^+}{p^+}\bar n^\mu + \frac{y l_T^\mu + q_T^\mu}{xyp^+}, \nonumber\\
 & t^\mu = \frac{2(1-y)x^2\tilde p^+ p^+}{y Q^2}\bar n^\mu +\frac{p^+}{\tilde p^+} n^\mu + \frac{2xyp^+}{Q^2} l_T^\mu. \label{f:tbart}
\end{align}
According to our definition, the transverse components of $l^\mu$ and $q^\mu$ are in the lepton plane which is just the $x-z$ plane. Therefore, $l_T=l_x=Q\sqrt{1-y}/y$ and $q_T=q_x=-Q\sqrt{1-y}$. Then the second term in the first line in Eq. (\ref{f:tbart}) vanishes and
\begin{align}
  p^+ \bar t^\mu =\frac{ (2-y)}{y}\tilde p^+ \bar n^\mu.
\end{align}
$\tilde p^+$ and $p^+$ do not have to be equal.

\begin{widetext}

\section{Twist-3 hadronic tensor} \label{sec:appendixa}

There are two origins of the twist-3 hadronic tensor, one from the quark-quark correlator and the other from the quark-gluon-quark correlator. We first consider contributions from the quark-quark correlator.
 Inserting the twist-3 TMD PDFs in  Eqs. (\ref{eq:Xi0Peven}), (\ref{eq:Xi0Podd}) and hadron part given in Eq. (\ref{f:hardamp}) into Eq. (\ref{f:hadronlead}),  we obtain
\begin{align}
  W^{\mu\nu}_{t3,q} &=\frac{ f^\perp}{p\cdot q}\bigg[ +c_1^q \left(k_T^{\{\mu} t^{\nu\}}q^- + k_T^{\{\mu} \bar t^{\nu\}}(1-y)xp^+ + k_T^{\{\mu} q_T^{\nu\}}- g^{\mu\nu}k_T\cdot q_T\right)   \nonumber\\
  &\quad \quad \quad   +ic_3^q\left(\tilde k_T^{[\mu} t^{\nu]}q^- -\tilde k_T^{[\mu} \bar t^{\nu]}(1-y)xp^+ - \bar t^{[\mu} t^{\nu]}\varepsilon_T^{ k q}\right) \bigg] \nonumber\\
  &+\frac{\lambda_h f^\perp_L}{p\cdot q}\bigg[- c_1^q \left(\tilde k_T^{\{\mu} t^{\nu\}}q^- + \tilde k_T^{\{\mu} \bar t^{\nu\}}(1-y)xp^+ + \tilde k_T^{\{\mu} q_T^{\nu\}}- g^{\mu\nu}\varepsilon_T^{ k q}\right) \nonumber\\
  &\quad \quad \quad +ic_3^q\left( k_T^{[\mu} t^{\nu]}q^- -k_T^{[\mu} \bar t^{\nu]}(1-y)xp^+ - \bar t^{[\mu} t^{\nu]}k_T\cdot q_T\right) \bigg] \nonumber\\
 &+\frac{ g^\perp }{p\cdot q}\bigg[- c_3^q \left(\tilde k_T^{\{\mu} t^{\nu\}}q^- + \tilde k_T^{\{\mu} \bar t^{\nu\}}(1-y)xp^+ + \tilde k_T^{\{\mu} q_T^{\nu\}}- g^{\mu\nu}\varepsilon_T^{ k q}\right)  \nonumber\\
  &\quad \quad \quad +ic_1^q\left( k_T^{[\mu} t^{\nu]}q^- -k_T^{[\mu} \bar t^{\nu]}(1-y)xp^+ - \bar t^{[\mu} t^{\nu]}k_T\cdot q_T\right) \bigg]\nonumber\\
 &+\frac{\lambda g^\perp_L}{p\cdot q}\bigg[ -c_3^q \left(k_T^{\{\mu} t^{\nu\}}q^- + k_T^{\{\mu} \bar t^{\nu\}}(1-y)xp^+ + k_T^{\{\mu} q_T^{\nu\}}- g^{\mu\nu}k_T\cdot q_T\right)   \nonumber\\
  &\quad \quad \quad   -ic_1^q\left(\tilde k_T^{[\mu} t^{\nu]}q^- -\tilde k_T^{[\mu} \bar t^{\nu]}(1-y)xp^+ - \bar t^{[\mu} t^{\nu]}\varepsilon_T^{ k q}\right) \bigg], \nonumber \\
  &+\frac{Mf_T}{p\cdot q}\Bigg\{-c_1^q \bigg[\tilde S_T^{\{\mu} t^{\nu\}}q^- + \tilde S_T^{\{\mu} \bar t^{\nu\}}(1-y)xp^+ + \tilde S_T^{\{\mu} q_T^{\nu\}}-  g^{\mu\nu} \varepsilon_T^{  q S}\bigg]  \nonumber\\
  &\quad \quad \quad+ ic_3^q \bigg[ S_T^{[\mu} t^{\nu]}q^- - S_T^{[\mu} \bar t^{\nu]}(1-y)xp^+ - \bar t^{[\mu} t^{\nu]}S_T\cdot q_T \bigg]\Bigg\}   \nonumber\\
   &-\frac{f_T^\perp}{p\cdot q} \Bigg\{+c_1^q \bigg[ k_T^{\{\mu} t^{\nu\}}q^- + k_T^{\{\mu} \bar t^{\nu\}}(1-y)xp^+ + k_T^{\{\mu} q_T^{\nu\}}-  g^{\mu\nu} k_T\cdot q_T \bigg]  \nonumber\\
  &\quad\quad \quad + ic_3^q \bigg[ \tilde k_T^{[\mu} t^{\nu]}q^- -\tilde k_T^{[\mu} \bar t^{\nu]}(1-y)xp^+ - \bar t^{[\mu} t^{\nu]}\varepsilon_T^{qk}\bigg] \Bigg\}\frac{\varepsilon_T^{kS}}{M}  \nonumber \\
  &-\frac{f_T^\perp}{p\cdot q} \Bigg\{-c_1^q \bigg[\tilde S_T^{\{\mu} t^{\nu\}}q^- + \tilde S_T^{\{\mu} \bar t^{\nu\}}(1-y)xp^+ + \tilde S_T^{\{\mu} q_T^{\nu\}}- g^{\mu\nu} \varepsilon_T^{  q S}\bigg]  \nonumber\\
  &\quad\quad \quad + ic_3^q \bigg[ S_T^{[\mu} t^{\nu]}q^- - S_T^{[\mu} \bar t^{\nu]}(1-y)xp^+ - \bar t^{[\mu} t^{\nu]}S_T\cdot q_T \bigg]\Bigg\} \frac{k_T^2}{2M}    \nonumber\\
   &+\frac{Mg_T}{p\cdot q} \Bigg\{-c_3^q \bigg[ S_T^{\{\mu} t^{\nu\}}q^- +  S_T^{\{\mu} \bar t^{\nu\}}(1-y)xp^+ +  S_T^{\{\mu} q_T^{\nu\}}- g^{\mu\nu}S_T\cdot q_T\bigg]  \nonumber\\
  & \quad\quad \quad- ic_1^q \bigg[ \tilde S_T^{[\mu} t^{\nu]}q^- -\tilde S_T^{[\mu} \bar t^{\nu]}(1-y)xp^+ - \bar t^{[\mu} t^{\nu]}\varepsilon_T^{  q S} \bigg]\Bigg\}   \nonumber\\
   & +\frac{g_T^\perp}{p\cdot q} \Bigg\{+c_3^q \bigg[ k_T^{\{\mu} t^{\nu\}}q^- + k_T^{\{\mu} \bar t^{\nu\}}(1-y)xp^+ + k_T^{\{\mu} q_T^{\nu\}}-  g^{\mu\nu} k_T\cdot q_T \bigg]  \nonumber\\
  &\quad\quad \quad + ic_1^q \bigg[ \tilde k_T^{[\mu} t^{\nu]}q^- -\tilde k_T^{[\mu} \bar t^{\nu]}(1-y)xp^+ - \bar t^{[\mu} t^{\nu]}\varepsilon_T^{qk}\bigg] \Bigg\}\frac{k_T\cdot S_T}{M}  \nonumber\\
  &-\frac{g_T^\perp}{p\cdot q} \Bigg\{+c_3^q \bigg[ S_T^{\{\mu} t^{\nu\}}q^- +  S_T^{\{\mu} \bar t^{\nu\}}(1-y)xp^+ +  S_T^{\{\mu} q_T^{\nu\}}-  g^{\mu\nu}S_T\cdot q_T\bigg]  \nonumber\\
  & \quad\quad \quad+ic_1^q \bigg[ \tilde S_T^{[\mu} t^{\nu]}q^- -\tilde S_T^{[\mu} \bar t^{\nu]}(1-y)xp^+ - \bar t^{[\mu} t^{\nu]}\varepsilon_T^{  q S} \bigg]\Bigg\}\frac{k_T^2}{2M}.
  \label{f:wt31}
\end{align}
The subscript $q$ denotes hadronic tensor from the quark-quark correlator. This expression does not satisfy the current conservation law, i. e.,  $q_\mu W^{\mu\nu}_{t3,q} \neq 0$ due to the incompleteness of the twist-3 hadronic tensor.

The quark-gluon-quark correlator comes from the one gluon-exchanging process. From the operator definition of the  hadronic tensor we have
\begin{align}
  W^{\mu\nu}_{t3, L}=\frac{1}{2p\cdot q} {\rm{Tr}} \left[\hat \Phi_\rho^{(1)}(x, k_T) \hat H^{\mu\nu,\rho}_{ZZ}(q, k_1, k_2) \right], \label{f:hadronnext}
\end{align}
where $L$ denotes the left-cut \cite{Liang:2006wp}, $\hat \Phi_\rho^{(1)}$ is the quark-gluon-quark correlator  given in Eq. (\ref{f:Phi1}).
$\hat H^{\mu\nu,\rho}_{ZZ}$ is the hard scattering amplitude,
\begin{align}
  \hat H^{\mu\nu, \rho}_{ZZ} =\Gamma^{\mu,q}\frac{\slashed k_2 +\slashed q}{(k_2+q)^2}\gamma^\rho(\slashed k_1+\slashed q) \Gamma^{\nu,q}. \label{f:hardamp1}
\end{align}
To simplify the hard scattering amplitude, here we use the approximation that only the plus component of the momenta exist in $\frac{\slashed k_2 +\slashed q}{(k_2+q)^2}$.  Under this  approximation, we can rewritten the hard scattering amplitude as
\begin{align}
  \hat H^{\mu\nu,\rho}_{ZZ} =\frac{1}{2q^-} \Gamma^{\mu,q}\slashed{\bar t} \gamma^\rho(\slashed k_1+\slashed q) \Gamma^{\nu,q}. \label{f:hardamp1plus}
\end{align}
Inserting Eqs. (\ref{eq:Xi1Peven}), (\ref{eq:Xi1Podd}) and  (\ref{f:hardamp1plus}) into (\ref{f:hadronnext}) and using  Eq. (\ref{f:harden}), we have
\begin{align}
  W^{\mu\nu}_{t3, L} = & - \left[c_1^q \left(\bar t^\mu \bar t^\nu \frac{k_T\cdot q_T}{q^-} - k_T^\nu \bar t^\mu\right) - ic_3^q  \left(\bar t^\mu \bar t^\nu \frac{\varepsilon_T^{qk}}{q^-} - \tilde k_T^\nu \bar t^\mu\right)  \right]\frac{xp^+}{p\cdot q} f^\perp \nonumber \\
  & +  \left[ic_1^q \left(\bar t^\mu \bar t^\nu \frac{k_T\cdot q_T}{q^-} - k_T^\nu \bar t^\mu\right) + c_3^q  \left(\bar t^\mu \bar t^\nu \frac{\varepsilon_T^{qk}}{q^-} - \tilde k_T^\nu \bar t^\mu\right)  \right]\frac{xp^+}{p\cdot q} g^\perp \nonumber \\
  &+ \left[ ic_3^q  \left(\bar t^\mu \bar t^\nu \frac{k_T\cdot q_T}{q^-} -  k_T^\nu \bar t^\mu\right)  +c_1^q \left(\bar t^\mu \bar t^\nu \frac{\varepsilon_T^{qk}}{q^-} - \tilde k_T^\nu \bar t^\mu\right) \right]\frac{xp^+}{p\cdot q} \lambda_h f_L^\perp \nonumber  \\
  & + \left[  c_3^q  \left(\bar t^\mu \bar t^\nu \frac{k_T\cdot q_T}{q^-} -  k_T^\nu \bar t^\mu\right) - i c_1^q \left(\bar t^\mu \bar t^\nu \frac{\varepsilon_T^{qk}}{q^-} - \tilde k_T^\nu \bar t^\mu\right) \right]\frac{xp^+}{p\cdot q} \lambda_h g_L^\perp \nonumber\\
  &+ \left[ ic_3^q  \left(\bar t^\mu \bar t^\nu \frac{S_T\cdot q_T}{q^-} -  S_T^\nu \bar t^\mu\right)  +c_1^q \left(\bar t^\mu \bar t^\nu \frac{\varepsilon_T^{qS}}{q^-} - \tilde S_T^\nu \bar t^\mu\right) \right]\frac{xp^+}{p\cdot q} M f_T \nonumber \\
 & - \left\{ - \left[c_1^q \left(\bar t^\mu \bar t^\nu \frac{k_T\cdot q_T}{q^-} - k_T^\nu \bar t^\mu\right) - ic_3^q  \left(\bar t^\mu \bar t^\nu \frac{\varepsilon_T^{qk}}{q^-} - \tilde k_T^\nu \bar t^\mu\right)  \right]\frac{\varepsilon_T^{kS}}{M} \right. \nonumber \\
 &\quad  \ \  \left. +\left[ ic_3^q  \left(\bar t^\mu \bar t^\nu \frac{S_T\cdot q_T}{q^-} -  S_T^\nu \bar t^\mu\right)  +c_1^q \left(\bar t^\mu \bar t^\nu \frac{\varepsilon_T^{qS}}{q^-} - \tilde S_T^\nu \bar t^\mu\right) \right]\frac{k_{T M}^2}{2M} \right\} \frac{xp^+}{p\cdot q}f_T^\perp \nonumber \\
  &+ \left[ c_3^q  \left(\bar t^\mu \bar t^\nu \frac{S_T\cdot q_T}{q^-} -  S_T^\nu \bar t^\mu\right)  - i c_1^q \left(\bar t^\mu \bar t^\nu \frac{\varepsilon_T^{qS}}{q^-} - \tilde S_T^\nu \bar t^\mu\right) \right]\frac{xp^+}{p\cdot q} M g_T \nonumber \\
 & + \left\{ - \left[  c_3^q  \left(\bar t^\mu \bar t^\nu \frac{k_T\cdot q_T}{q^-} -  k_T^\nu \bar t^\mu\right) - i c_1^q \left(\bar t^\mu \bar t^\nu \frac{\varepsilon_T^{qk}}{q^-} - \tilde k_T^\nu \bar t^\mu\right) \right] \frac{k_T\cdot S_T}{M} \right. \nonumber \\
 &\quad \ \  \left. + \left[ c_3^q  \left(\bar t^\mu \bar t^\nu \frac{S_T\cdot q_T}{q^-} -  S_T^\nu \bar t^\mu\right)  - i c_1^q \left(\bar t^\mu \bar t^\nu \frac{\varepsilon_T^{qS}}{q^-} - \tilde S_T^\nu \bar t^\mu\right) \right]\frac{k_{T M}^2}{2M} \right\}\frac{xp^+}{p\cdot q} g_T^\perp.   \label{f:wt3l}
\end{align}
Subscript $L$ denotes the left-cut tensor. We note that TMD PDFs marked with subscript $d$ have been reexpressed in terms of TMD PDFs without subscript by using relation \cite{Chen:2020ugq}
\begin{align}
f_{d S}^{K}-g_{d S}^{K}=-x\left(f_{S}^{K}-i g_{S}^{K}\right),\label{f:formulaEOM}
\end{align}
where $K$ can be $\perp$ and $S$ can be $L$ and $T$. It straightforward to obtain the result above. For example, from Eqs. (\ref{f:hadronnext})-(\ref{f:hardamp1plus}), we calculate the trace and obtain
\begin{align}
  W^{\mu\nu}_{t3, L} =\left[c_1^q \left( \bar t^\mu \bar t^\nu \frac{k_T \cdot q_T}{q^-}- k_T^\nu \bar t^\mu \right) + ic_3^q \left(\tilde{k}^\nu_T \bar t^\mu - \bar t^\mu \bar t^\nu \frac{\varepsilon_T^{qk}}{q^-}\right)\right]\frac{p^+}{p\cdot q}(f^\perp_d -g^\perp_d). \label{f:wmunufgd}
\end{align}
as long as we only consider $f^\perp$ and $g^\perp$ terms.
Substituting Eq. (\ref{f:formulaEOM}) into Eq. (\ref{f:wmunufgd}) gives
\begin{align}
  W^{\mu\nu}_{t3, L} =\left[c_1^q \left( \bar t^\mu \bar t^\nu \frac{k_T \cdot q_T}{q^-}- k_T^\nu \bar t^\mu \right) + ic_3^q \left(\tilde{k}^\nu_T \bar t^\mu - \bar t^\mu \bar t^\nu \frac{\varepsilon_T^{qk}}{q^-}\right)\right]\frac{-xp^+}{p\cdot q}(f^\perp -ig^\perp), \label{f:wmunufg}
\end{align}
which corresponds to the first and second lines in Eq. (\ref{f:wt3l}). In order to obtain the complete result of the twist-3 hadronic tensor from the quark-gluon-quark correlator, we sum the left-cut and the right-cut terms together and obtain
\begin{align}
    W^{\mu\nu}_{t3, L}+  W^{\mu\nu}_{t3, R}= & - \left[c_1^q \left(\bar t^\mu \bar t^\nu \frac{2x p^+}{q^-} k_T\cdot q_T - xp^+ k_T^{\{\mu} \bar t^{\nu\}}\right) - ic_3^q  x p^+\tilde k_T^{[\mu} \bar t^{\nu]} \right]\frac{1}{p\cdot q} f^\perp \nonumber \\
  & +\left[c_3^q  \left(\bar t^\mu \bar t^\nu \frac{2x p^+}{q^-}\varepsilon_T^{kq} - xp^+ \tilde  k_T^{\{\mu} \bar t^{\nu\}}\right)+ic_1^qx p^+ k_T^{[\mu} \bar t^{\nu]}   \right]\frac{1}{p\cdot q} g^\perp \nonumber \\
  & +\left[c_1^q  \left(\bar t^\mu \bar t^\nu \frac{2x p^+}{q^-}\varepsilon_T^{kq} - xp^+ \tilde  k_T^{\{\mu} \bar t^{\nu\}}\right)+ic_3^q x p^+k_T^{[\mu} \bar t^{\nu]}   \right] \frac{1}{p\cdot q} \lambda_h f_L^\perp \nonumber  \\
  & + \left[c_3^q \left(\bar t^\mu \bar t^\nu \frac{2x p^+}{q^-} k_T\cdot q_T - xp^+ k_T^{\{\mu} \bar t^{\nu\}}\right) - ic_1^q x p^+ \tilde k_T^{[\mu} \bar t^{\nu]} \right]\frac{1}{p\cdot q} \lambda_h g_L^\perp \nonumber\\
  &+ \left[ c_1^q \left(\bar t^\mu \bar t^\nu \frac{2xp^+}{q^-}\varepsilon_T^{qS} - \tilde S_T^{\{\mu} \bar t^{\nu\}} \right)+ ic_3^q  x p^+ S_T^{[\mu} \bar t^{\nu]}\right]\frac{1}{p\cdot q} M f_T \nonumber \\
 & - \left\{ - \left[c_1^q \left(\bar t^\mu \bar t^\nu \frac{2x p^+}{q^-} k_T\cdot q_T - xp^+ k_T^{\{\mu} \bar t^{\nu\}}\right) - ic_3^q  x p^+\tilde k_T^{[\mu} \bar t^{\nu]}   \right]\frac{\varepsilon_T^{kS}}{M} \right. \nonumber \\
 &\quad  \ \  \left. +\left[ c_1^q \left(\bar t^\mu \bar t^\nu \frac{2xp^+}{q^-}\varepsilon_T^{qS} - \tilde S_T^{\{\mu} \bar t^{\nu\}} \right)+ ic_3^q  x p^+ S_T^{[\mu} \bar t^{\nu]} \right]\frac{k_{T M}^2}{2M} \right\} \frac{1}{p\cdot q}f_T^\perp \nonumber \\
  &+ \left[ c_3^q  \left(\bar t^\mu \bar t^\nu \frac{2xp^+}{q^-} S_T\cdot q_T-  xp^+S_T^{\{\mu} \bar t^{\nu\}}\right)  - i c_1^qxp^+ \tilde S_T^{[\mu} \bar t^{\nu]} \right]\frac{1}{p\cdot q} M g_T \nonumber \\
 & + \left\{ - \left[  c_3^q  \left(\bar t^\mu \bar t^\nu \frac{2xp^+}{q^-}k_T\cdot q_T -  k_T^{\{\mu} \bar t^{\mu\}}\right) - i c_1^q  \tilde k_T^{[\mu} \bar t^{\nu]}\right] \frac{k_T\cdot S_T}{M} \right. \nonumber \\
 &\quad \ \  \left. + \left[ c_3^q  \left(\bar t^\mu \bar t^\nu \frac{2xp^+}{q^-} S_T\cdot q_T-  xp^+S_T^{\{\mu} \bar t^{\nu\}}\right)  - i c_1^qxp^+ \tilde S_T^{[\mu} \bar t^{\nu]} \right]\frac{k_{T M}^2}{2M} \right\}\frac{1}{p\cdot q} g_T^\perp.     \label{f:wt3lr}
\end{align}
One knows the relationship that $ W^{\mu\nu}_{t3, L}= (W^{\nu\mu}_{t3, R})^*$.
Finally, we sum Eqs. (\ref{f:wt31}) and (\ref{f:wt3lr}) to obtain the complete hadronic tensor given in Eq. (\ref{f:wt3}).
\end{widetext}

\end{appendix}


\begin{thebibliography}{0}

\bibitem{Accardi:2012qut}
  A.~Accardi {\it et al.},
  Eur.\ Phys.\ J.\ A {\bf 52}, no. 9, 268 (2016)
  doi:10.1140/epja/i2016-16268-9
  [arXiv:1212.1701 [nucl-ex]].

\bibitem{AbdulKhalek:2021gbh}
R.~Abdul Khalek, A.~Accardi, J.~Adam, D.~Adamiak, W.~Akers, M.~Albaladejo, A.~Al-bataineh, M.~G.~Alexeev, F.~Ameli and P.~Antonioli, \textit{et al.}
[arXiv:2103.05419 [physics.ins-det]].

\bibitem{Anderle:2021wcy}
D.~P.~Anderle, V.~Bertone, X.~Cao, L.~Chang, N.~Chang, G.~Chen, X.~Chen, Z.~Chen, Z.~Cui and L.~Dai, \textit{et al.}
Front. Phys. (Beijing) \textbf{16}, no.6, 64701 (2021)
doi:10.1007/s11467-021-1062-0
[arXiv:2102.09222 [nucl-ex]].



\bibitem{Mulders:1995dh}
  P.~J.~Mulders and R.~D.~Tangerman,
  Nucl.\ Phys.\ B {\bf 461}, 197 (1996)
  Erratum: [Nucl.\ Phys.\ B {\bf 484}, 538 (1997)]
  doi:10.1016/S0550-3213(96)00648-7, 10.1016/0550-3213(95)00632-X
  [hep-ph/9510301].


\bibitem{Bacchetta:2006tn}
  A.~Bacchetta, M.~Diehl, K.~Goeke, A.~Metz, P.~J.~Mulders and M.~Schlegel,
  JHEP {\bf 0702}, 093 (2007)
  doi:10.1088/1126-6708/2007/02/093
  [hep-ph/0611265].


\bibitem{Beneke:2022obx}
M.~Beneke, M.~Garny, S.~Jaskiewicz, J.~Strohm, R.~Szafron, L.~Vernazza and J.~Wang,
JHEP \textbf{07}, 144 (2022)
doi:10.1007/JHEP07(2022)144
[arXiv:2205.04479 [hep-ph]].


\bibitem{Gamberg:2022lju}
L.~Gamberg, Z.~B.~Kang, D.~Y.~Shao, J.~Terry and F.~Zhao,
[arXiv:2211.13209 [hep-ph]].


\bibitem{Rodini:2023plb}
S.~Rodini and A.~Vladimirov,
[arXiv:2306.09495 [hep-ph]].

\bibitem{Boer:1997nt}
D.~Boer and P.~J.~Mulders,
Phys. Rev. D \textbf{57}, 5780-5786 (1998)
doi:10.1103/PhysRevD.57.5780
[arXiv:hep-ph/9711485 [hep-ph]].




\bibitem{Accardi:2017pmi}
A.~Accardi and A.~Bacchetta,
Phys. Lett. B \textbf{773}, 632-638 (2017)
doi:10.1016/j.physletb.2017.08.074
[arXiv:1706.02000 [hep-ph]].

\bibitem{Accardi:2019luo}
A.~Accardi and A.~Signori,
Phys. Lett. B \textbf{798}, 134993 (2019)
doi:10.1016/j.physletb.2019.134993
[arXiv:1903.04458 [hep-ph]].

\bibitem{Accardi:2020iqn}
A.~Accardi and A.~Signori,
Eur. Phys. J. C \textbf{80}, no.9, 825 (2020)
doi:10.1140/epjc/s10052-020-8380-1
[arXiv:2005.11310 [hep-ph]].

\bibitem{Kang:2020xyq}
Z.~B.~Kang, K.~Lee and F.~Zhao,
Phys. Lett. B \textbf{809}, 135756 (2020)
doi:10.1016/j.physletb.2020.135756
[arXiv:2005.02398 [hep-ph]].


\bibitem{Liang:2006wp}
  Z.~t.~Liang and X.~N.~Wang,
  Phys.\ Rev.\ D {\bf 75}, 094002 (2007)
  [hep-ph/0609225].




\bibitem{Song:2010pf}
 Y.~k.~Song, J.~h.~Gao, Z.~t.~Liang and X.~N.~Wang,
  Phys.\ Rev.\ D {\bf 83}, 054010 (2011)
  doi:10.1103/PhysRevD.83.054010
  [arXiv:1012.4179 [hep-ph]].


\bibitem{Song:2013sja}
  Y.~k.~Song, J.~h.~Gao, Z.~t.~Liang and X.~N.~Wang,
  Phys.\ Rev.\ D {\bf 89}, no. 1, 014005 (2014)
  doi:10.1103/PhysRevD.89.014005
  [arXiv:1308.1159 [hep-ph]].


\bibitem{Wei:2016far}
  S.~y.~Wei, Y.~k.~Song, K.~b.~Chen and Z.~t.~Liang,
  Phys.\ Rev.\ D {\bf 95}, no. 7, 074017 (2017)
  doi:10.1103/PhysRevD.95.074017
  [arXiv:1611.08688 [hep-ph]].



\bibitem{Gutierrez-Reyes:2018qez}
D.~Gutierrez-Reyes, I.~Scimemi, W.~J.~Waalewijn and L.~Zoppi,
Phys. Rev. Lett. \textbf{121}, no.16, 162001 (2018)
doi:10.1103/PhysRevLett.121.162001
[arXiv:1807.07573 [hep-ph]].


 \bibitem{Gutierrez-Reyes:2019vbx}
D.~Gutierrez-Reyes, I.~Scimemi, W.~J.~Waalewijn and L.~Zoppi,
JHEP \textbf{10}, 031 (2019)
doi:10.1007/JHEP10(2019)031
[arXiv:1904.04259 [hep-ph]].


\bibitem{Kang:2020fka}
Z.~B.~Kang, X.~Liu, S.~Mantry and D.~Y.~Shao,
Phys. Rev. Lett. \textbf{125}, 242003 (2020)
doi:10.1103/PhysRevLett.125.242003
[arXiv:2008.00655 [hep-ph]].


\bibitem{Arratia:2020ssx}
M.~Arratia, Y.~Makris, D.~Neill, F.~Ringer and N.~Sato,
Phys. Rev. D \textbf{104}, no.3, 034005 (2021)
doi:10.1103/PhysRevD.104.034005
[arXiv:2006.10751 [hep-ph]].



\bibitem{Chen:2020ugq}
K.~B.~Chen and W.~H.~Yang,
Phys. Rev. D \textbf{101}, no.9, 096017 (2020)
doi:10.1103/PhysRevD.101.096017
[arXiv:2004.01359 [hep-ph]].

\bibitem{Yang:2020qsk}
W.~Yang,
Phys. Rev. D \textbf{103}, no.1, 016011 (2021)
doi:10.1103/PhysRevD.103.016011
[arXiv:2011.10212 [hep-ph]].

\bibitem{Yang:2022xwy}
W.~Yang and X.~Yang,
Phys. Rev. D \textbf{106}, no.9, 093003 (2022)
doi:10.1103/PhysRevD.106.093003
[arXiv:2209.01629 [hep-ph]].




\bibitem{Liu:2020dct}
X.~Liu, F.~Ringer, W.~Vogelsang and F.~Yuan,
Phys. Rev. D \textbf{102}, no.9, 094022 (2020)
doi:10.1103/PhysRevD.102.094022
[arXiv:2007.12866 [hep-ph]].


\bibitem{Liu:2018trl}
X.~Liu, F.~Ringer, W.~Vogelsang and F.~Yuan,
Phys. Rev. Lett. \textbf{122}, no.19, 192003 (2019)
doi:10.1103/PhysRevLett.122.192003
[arXiv:1812.08077 [hep-ph]].





\bibitem{Kang:2021ffh}
Z.~B.~Kang, K.~Lee, D.~Y.~Shao and F.~Zhao,
JHEP \textbf{11}, 005 (2021)
doi:10.1007/JHEP11(2021)005
[arXiv:2106.15624 [hep-ph]].


\bibitem{Kang:2022dpx}
Z.~B.~Kang, K.~Lee, D.~Y.~Shao and F.~Zhao,
JPS Conf. Proc. \textbf{37}, 020128 (2022)
doi:10.7566/JPSCP.37.020128
[arXiv:2201.04582 [hep-ph]].



\bibitem{H1:2021wkz}
V.~Andreev \textit{et al.} [H1],
Phys. Rev. Lett. \textbf{128}, no.13, 132002 (2022)
doi:10.1103/PhysRevLett.128.132002
[arXiv:2108.12376 [hep-ex]].




\bibitem{Yang:2023vyv}
W.~Yang and X.~Yang,
Nucl. Phys. B \textbf{990}, 116181 (2023)
doi:10.1016/j.nuclphysb.2023.116181



\bibitem{Cahn:1977uu}
  R.~N.~Cahn and F.~J.~Gilman,
  Phys.\ Rev.\ D {\bf 17}, 1313 (1978).
  doi:10.1103/PhysRevD.17.1313.

\bibitem{Mulders0}
P.~J.~Mulders,
``Transverse momentum dependence in structure functions in hard scattering processes,''
\url{http://www.nat.vu.nl/~mulders/correlations-0.pdf} (unpublished).



\bibitem{Sivers:1989cc}
D.~W.~Sivers,
Phys. Rev. D \textbf{41}, 83 (1990)
doi:10.1103/PhysRevD.41.83

\bibitem{Sivers:1990fh}
D.~W.~Sivers,
Phys. Rev. D \textbf{43}, 261-263 (1991)
doi:10.1103/PhysRevD.43.261


\bibitem{Schmidt:2015zda}
C.~Schmidt, J.~Pumplin, D.~Stump and C.~P.~Yuan,
Phys. Rev. D \textbf{93}, no.11, 114015 (2016)
doi:10.1103/PhysRevD.93.114015
[arXiv:1509.02905 [hep-ph]].









\bibitem{Anselmino:2005nn}
M.~Anselmino, M.~Boglione, U.~D'Alesio, A.~Kotzinian, F.~Murgia and A.~Prokudin,
Phys. Rev. D \textbf{71}, 074006 (2005)
doi:10.1103/PhysRevD.71.074006
[arXiv:hep-ph/0501196 [hep-ph]].


\bibitem{Signori:2013mda}
A.~Signori, A.~Bacchetta, M.~Radici and G.~Schnell,
JHEP \textbf{11}, 194 (2013)
doi:10.1007/JHEP11(2013)194
[arXiv:1309.3507 [hep-ph]].

\bibitem{Anselmino:2013lza}
M.~Anselmino, M.~Boglione, J.~O.~Gonzalez Hernandez, S.~Melis and A.~Prokudin,
JHEP \textbf{04}, 005 (2014)
doi:10.1007/JHEP04(2014)005
[arXiv:1312.6261 [hep-ph]].


\bibitem{Cammarota:2020qcw}
J.~Cammarota \textit{et al.} [Jefferson Lab Angular Momentum],
Phys. Rev. D \textbf{102}, no.5, 054002 (2020)
doi:10.1103/PhysRevD.102.054002
[arXiv:2002.08384 [hep-ph]].


\bibitem{Bacchetta:2022awv}
A.~Bacchetta, V.~Bertone, C.~Bissolotti, G.~Bozzi, M.~Cerutti, F.~Piacenza, M.~Radici and A.~Signori,
[arXiv:2206.07598 [hep-ph]].



\end{thebibliography}
\end{document}